\documentclass[english,prb, preprint]{revtex4-1}
\usepackage[T1]{fontenc}
\usepackage[latin9]{inputenc}
\usepackage{xcolor}
\usepackage{pdfcolmk}
\usepackage{amsmath}
\usepackage{amssymb}
\usepackage{graphicx}
\PassOptionsToPackage{normalem}{ulem}
\usepackage{ulem}

\makeatletter

\providecolor{lyxadded}{rgb}{0,0,1}
\providecolor{lyxdeleted}{rgb}{1,0,0}
\DeclareRobustCommand{\lyxsout}[1]{\ifx\\#1\else\sout{#1}\fi}

\usepackage{color}
\usepackage{amsmath}

\makeatother

\usepackage{babel}
\begin{document}
\title{Magnetic circular dichroism in hard x-ray Raman scattering as a probe
of local spin polarization}
\author{Manabu Takahashi}
\affiliation{Faculty of Engineering, Gunma University, Kiryu, Gunma 376, Japan}
\author{Nozomu Hiraoka}
\affiliation{National Synchrotron Radiation Research Center, Hsinchu 30076, Taiwan}
\begin{abstract}
We argue that the magnetic circular dichroism (MCD) of the hard x-ray
Raman scattering (XRS) could be used as an element selective probe
of local spin polarization. The magnitude of the XRS-MCD signal is
directly proportional to the local spin polarization when the angle
between the incident wavevector and the magnetization vector is $135^{\circ}$
or $-45^{\circ}$ . By comparing the experimental observation and
the configuration interaction calculation at the $L_{2,3}$ and $M_{2,3}$
edges of ferromagnetic iron, we suggest that the integrated MCD signal
in terms of the transferred energy could be used to estimate the local
spin moment even in the case where the application of the spin sum-rule
in X-ray absorption is questionable. We also point out that XRS-MCD
signal could be observed at the $M_{1}$ edge with a magnitude comparable
to that at the $M_{2,3}$ edge, although the spin-orbit coupling is
absent in the core orbital. By combining the XRS-MCD at various edges,
spin polarization distribution depending on the orbital magnetic quantum
number would be determined.
\end{abstract}
\maketitle

\section{Introduction}

X-ray magnetic circular dichroism (MCD) has been one of the powerful
tools to investigate the electronic structure in magnetic materials.
Particularly, owing to the orbital and spin sum-rules,\citep{Carra_1993,Thole_1992}
the MCD in the soft x-ray absorption spectroscopy (XAS) has been playing
crucial roles for elucidating the electronic structure at and around
the absorption site.\citep{Nakamura_2013} The MCD measurements in
x-ray emission and resonant inelastic scattering are also important
tools to clarify the electronic excitations in magnetic materials.\citep{Kotani_2001}

Recently, the MCD in the hard x-ray Raman scattering (XRS) at the
Fe $\mathrm{L}_{2,3}$-edges in the ferromagnetic iron has been investigated.\citep{Hiraoka_2015,Takahashi_2015}
The XRS is a kind of non-resonant inelastic x-ray scattering.\citep{Rueff_2010}
In the process of photon scattering, where an incident photon of energy
$\hbar\omega_{\mathrm{i}}$ is absorbed and a photon of energy $\hbar\omega_{\mathrm{f}}$
is emitted, the electron system in the initial ground state of the
energy $E_{\mathrm{i}}$ is excited to the final state of the energy
$E_{\mathrm{f}}=E_{\mathrm{i}}+\hbar\omega_{\mathrm{i}}-\hbar\omega_{\mathrm{f}}$.
The final state of the XRS is essentially the same with that of XAS:
A core hole is left behind at the scattering site and an electron
is added to the valence or conduction state. Therefore, the XRS intensity
as a function of transferred energy is similar to the soft x-ray absorption
coefficient as a function of the incident photon energy. Contrasting
to the XAS, the hard-in-hard-out feature of the XRS is preferable
for bulk sensitive measurements or the measurements under extreme
conditions. In addition, the XRS can access the final states that
are inaccessible by the dipole transition, because the non-dipole
transition matrix elements become significant for shallow core excitation.
By virtue of these features, the inner-core-exciting XRS has been
demonstrating its usefulness particularly to unveil the electronic
state of materials under extreme conditions.\citep{Sternemann_2016}
Recently, the electronic state of Fe in $\mathrm{Fe_{2}SiO_{4}}$,
$\mathrm{Fe_{2}O_{3}}$ and $\mathrm{FeS}$ under high pressure is
discussed by analyzing the XRS spectra at the Fe $M_{2,3}$ edge,
and a spin transition is revealed from the change of spectral curves
in $\mathrm{FeS}$.\citep{Nyrow_2014,Nyrow_2014_2} In addition, XRS
at the rare earth $N$ and $O$ edges has been extensively discussed.\citep{van_der_Laan_2012,Huotari_2015}

The XAS-MCD measurements are carried out in order to elucidate the
orbital and spin magnetic moment at the selected magnetic ion. In
the analysis of the MCD signal, the orbital- and the spin moment sum
rules play central roles. However, it is also known that the spin
sum rule has some limitations. The core hole level $j=l\pm1/2$ should
be clearly separated for safe application of the sum rule. Teramura
et al. showed that the deviation of the rule could amount to 30\%
for $\mathrm{Mn}^{2+}$ and reached at 230\% for Sm.\citep{Teramura1996,Teramura_1996}
The XAS-MCD signal can be observed also at the $M_{2,3}$ edge of
transition metals.\citep{Yoshida_1991,Koide_1991} However, it is
quite difficult to obtain the information about the local spin moment
from the observed MCD signal alone, because it is quite hard to apply
the spin sum rule due to the smallness of the spin orbit coupling
(SOC) of the 3p hole, the strong 3p-3d Coulomb interaction, and the
remarkable super-Coster-Kroning decay.\citep{Coster_1935} The MCD
signal at the $\mathrm{K}$-edge of transition metals also have been
observed. While it is bulk sensitive, we can only indirectly obtain
the information about orbital moment of the 3d state through the interaction
between the 3d and 4p states.\citep{Igarashi_1994,Igarashi_1996,Brouder_1996}
It is worth noting that the spin sum rule\citep{Thole_1993,van_der_Laan_1993}
of the X-ray photoemission spectroscopy, in which both of the photon
polarization and the spin of the emitted electron are exploited and
the magnetic dichroism is observed in the emission of electrons, can
be applied to estimate the ground state spin moment without the aforementioned
shortcomings of the XAS-MCD spin sum rule.

The XRS-MCD signals have been observed at the $L_{2,3}$ edges of
ferromagnetic iron by Hiraoka et al. \citep{Hiraoka_2015} The observe
MCD spectral curves as a function of the transferred energy depend
strongly on the angle $\alpha_{\mathrm{M}}$ (see fig. \ref{figscattering_geometry}.),
which is similar to the XAS-MCD spectral curve at the angle $\alpha_{\mathrm{M}}\sim0^{\circ}$
. In the previous paper,\citep{Takahashi_2015} we analyzed the XRS-MCD
signals within a one-electron theory and discussed the relation between
the spectral shape of the MCD signal and the angle $\alpha_{\mathrm{M}}$.
We elucidated that the XRS-MCD signals can be considered as a result
of the interference of the scattering amplitude due to the charge
transition with that due to the electric, the orbital magnetic, and
the spin magnetic transitions; their effects differently depend on
the angle $\alpha_{\mathrm{M}}$. Particularly, at $\alpha_{\mathrm{M}}=135^{\circ}$
or $-45^{\circ}$, the magnitude of the XRS-MCD signal is proportional
only to the local spin polarization. Therefore, the integrated XRS-MCD
signal could be also used as a probe of local spin moment.

The magnetic Compton scattering (MCS) technique,\citep{Cooper2004}
which reveals the distribution of the spin magnetic moment in the
momentum space, and the x-ray magnetic diffraction (XMD)\citep{Blume_1985}
also owes to the MCD effect. The mechanism causing the MCD in MCS
and XMD is different from that in XAS. The interaction between the
magnetic field of radiation and the electron spin and/or orbital magnetic
moments brings about the MCD effects in MCS and XMD. On the other
hand, in the XAS, the spin-orbit coupling in the inner shell plays
essential roles in provoking the MCD signal, since the electric field
of the radiation does not directly couple to the orbital and spin
magnetic moments. In the XRS, both mechanisms can induce the MCD signal
with the magnitude comparable to each other. In contrast to the MCS,
XRS-MCD may have advantageous features: the element selectivity and
the selection rules in transition process. At $\alpha_{\mathrm{M}}=135^{\circ}$,
the interference of scattering amplitude due to the charge- and spin-transitions
alone produces the MCD signal. Therefore, it is expected that the
XRS-MCD measurement could be used as a probe of local spin polarization
at the scattering site even if the SOC is absent.

The interaction between the 3p electrons and the 3d electrons is so
large that the M-edge excitation spectrum is expected to sensitively
reflect the 3d state. Besides the XRS or XAS, the M-edge excitation
of transition metal has been well measured using Electron Energy Loss
Spectroscopy (EELS), or $K\beta$ x-ray emission spectroscopy (XES).
Utilizing the surface sensitivity, the EELS is used to study the electronic
structure in thin films.\citep{Garvie_2004} The signal of the $K\beta$
XES is bulk sensitive, and the spin dependent spectrum caused by the
3p-3d exchange interaction is useful to elucidate the 3d state.\citep{Taguchi_1997}
The dichroic effect of the Fe $K\alpha_{1}$ emission spectrum has
also recently been observed.\citep{Inami_2017} In addition to these
techniques, XRS-MCD may become a useful technique to understand the
electronic structures under extreme conditions with exploiting its
bulk sensitivity, element and orbital selectivity.

In the next section, we briefly describe the XRS-MCD formula. The
model used to simulate the electronic state at the scattering site
is described in Section III. In Section IV we discuss the $L_{2,3}$
and $M_{2,3}$ edge XRS spectra by comparing the calculations and
the observations. The XRS spectra at the $M_{1}$ edge is also demonstrated.
The last section is devoted to the concluding remarks. Demonstrations
of the XRS-MCD spin sum rule are involved in the last section.

\section{Scattering intensity and MCD signal}

We assume that the electronic state is excited from the initial state
$\Phi_{\mathrm{i}}$ with energy $E_{\mathrm{i}}$ to the final state
$\Phi_{\mathrm{f}}$ with energy $E_{\mathrm{f}}$ by absorbing an
incident photon of polarization $\boldsymbol{e}_{\mathrm{i}}$, wave
vector $\boldsymbol{q}_{\mathrm{i}}$, and energy $\hbar\omega_{\mathrm{i}}$and
emitting a photon of polarization $\boldsymbol{e}_{\mathrm{f}}$,
wave vector $\boldsymbol{q}_{\mathrm{f}}$, and energy of $\hbar\omega_{\mathrm{f}}$.
In the final state, a core hole is left behind at the scattering site
and an electron is added to the valence or conduction state. The scattering
intensity may be proportional to the factor $\sum_{\Phi_{\text{f}}}\left|\bigl\langle\Phi_{\text{f}}|\sum_{i}\hat{f}(\boldsymbol{x}_{i})|\Phi_{\text{i}}\bigr\rangle\right|^{2}\delta_{E}$.
Here, $\delta_{E}$ represents the energy conservation delta function
$\delta\left(\Delta E+E_{\mathrm{i}}-E_{\mathrm{f}}\right)$ with
$\Delta E=\hbar\omega_{\mathrm{i}}-\hbar\omega_{\mathrm{f}}$; the
operator $\hat{f}$ is approximately given by the sum of the charge,
electric, orbital magnetic, and spin magnetic transition operators
$\hat{f}_{\mathrm{C}}$, $\hat{f}_{\mathrm{E}}$, $\hat{f}_{\mathrm{O}}$,
and $\hat{f}_{\mathrm{S}}$, which are given in equations (\ref{eq:fc}-\ref{eq:fs});\citep{Takahashi_2015}
$\boldsymbol{x}_{i}$ refers to the position $\boldsymbol{r}_{i}$
and spin $\boldsymbol{s}_{i}$ operators of the $i$th electron. The
transition operators $\hat{f}_{\mathrm{C}}$, $\hat{f}_{\mathrm{E}}$,
$\hat{f}_{\mathrm{O}}$, and $\hat{f}_{\mathrm{S}}$ are derived as
the first- and second-order perturbation in terms of the interaction
between electrons and electromagnetic field in the non-relativistic
Hamiltonian.\citep{Fr_hlich_1993} The perturbation terms of the higher
order than $\left(\hbar\omega_{\mathrm{i}}/m_{\mathrm{e}}c^{2}\right)^{2}$
may be safely ignored, where $m_{\mathrm{e}}c^{2}$ is the electron
rest energy. To handle the second-order perturbation terms, we assume
that a core electron is excited to form an intermediate state $\Phi_{n}$
and the electron successively comes down to an energy level near the
lowest unoccupied state to form a final electronic state $\Phi_{\mathrm{f}}$,
and take the non-resonant limit, in which we ignore the energy difference
between the intermediate electronic state energy $E_{n}$ and the
initial electronic state energy $E_{\mathrm{i}}$ in the energy denominator
assuming $E_{n}-E_{\mathrm{i}}\ll\hbar\omega_{\mathrm{i}},\ \hbar\omega_{\mathrm{f}}$.
Then, we exploit the completeness of the intermediate electronic state
$\Phi_{n}$, and neglect the terms of the order $1-\omega_{\mathrm{f}}/\omega_{\mathrm{i}}$.
Thus, the transition operators may be obtained as \begin{subequations}
\begin{align}
 & \hat{f}{}_{\mathrm{C}}\left(\boldsymbol{x}\right)=\boldsymbol{e}_{\text{f}}\cdot\boldsymbol{e}_{\text{i}}e^{i\boldsymbol{Q}\cdot\boldsymbol{r}},\label{eq:fc}\\
 & \hat{f}{}_{\mathrm{E}}\left(\boldsymbol{x}\right)=\frac{i\Delta E}{\alpha m_{\mathrm{e}}c^{2}}\boldsymbol{A}\cdot\boldsymbol{G}\left(\boldsymbol{Q},\boldsymbol{r}\right),\label{eq:fe}\\
 & \hat{f}{}_{\mathrm{O}}\left(\boldsymbol{x}\right)=-\frac{iE_{Q}}{2m_{\mathrm{e}}c^{2}}\boldsymbol{A}\cdot\hat{\boldsymbol{Q}}\times\boldsymbol{L}\left(\boldsymbol{Q},\boldsymbol{r}\right),\label{eq:fo}\\
 & \hat{f}{}_{\mathrm{S}}\left(\boldsymbol{x}\right)=-\frac{i\overline{E}}{2m_{\mathrm{e}}c^{2}}\boldsymbol{h}\cdot\boldsymbol{\sigma}e^{i\boldsymbol{Q}\cdot\boldsymbol{r}},\label{eq:fs}
\end{align}
\end{subequations}where $\boldsymbol{\sigma}$, $\boldsymbol{Q}$,
$\boldsymbol{A},$and $\boldsymbol{h}$ are defined as $\boldsymbol{\sigma}=2\boldsymbol{s}/\hbar$,
$\boldsymbol{Q}=\boldsymbol{q}_{\mathrm{i}}-\boldsymbol{q}_{\mathrm{f}}$,
$\boldsymbol{A}=\left(\boldsymbol{e}_{\text{f}}\cdot\hat{\boldsymbol{q}}_{\text{i}}\right)\boldsymbol{e}_{\text{i}}+\left(\boldsymbol{e}_{\text{i}}\cdot\hat{\boldsymbol{q}}_{\text{f}}\right)\boldsymbol{e}_{\text{f}}$,
and\textbf{ $\boldsymbol{h}=\frac{\hbar\omega_{\text{f}}}{\overline{E}}\left(\boldsymbol{e}_{\text{i}}\cdot\hat{\boldsymbol{q}}_{\text{f}}\right)\left(\hat{\boldsymbol{q}}_{\text{f}}\times\boldsymbol{e}_{\text{f}}\right)-\frac{\hbar\omega_{\text{i}}}{\overline{E}}\left(\boldsymbol{e}_{\text{f}}\cdot\hat{\boldsymbol{q}}_{\text{i}}\right)\left(\hat{\boldsymbol{q}}_{\text{i}}\times\boldsymbol{e}_{\text{i}}\right)+\boldsymbol{e}_{\text{f}}\times\boldsymbol{e}_{\text{i}}-\left(\hat{\boldsymbol{q}}_{\text{f}}\times\boldsymbol{e}_{\text{f}}\right)\times\left(\hat{\boldsymbol{q}}_{\text{i}}\times\boldsymbol{e}_{\text{i}}\right)$};
$\hat{\boldsymbol{q}}_{\mathrm{i(f)}}$ and $\hat{\boldsymbol{Q}}$
are the unit vectors $\boldsymbol{q}_{\mathrm{i(f)}}/q_{\mathrm{i(f)}}$
and $\boldsymbol{Q}/Q$, respectively. $E_{Q}$, and $\bar{E}$ are
energies defined as $E_{Q}=\hbar cQ,$ and $\bar{E}=\hbar\left(\omega_{\mathrm{i}}+\omega_{\mathrm{f}}\right)/2$,
respectively. $\alpha$ is the fine structure constant. The operators
$\hat{f}_{\mathrm{E}}$ and $\hat{f}_{\mathrm{O}}$ are deduced from
the terms including the linear momentum operator $\boldsymbol{p}=-i\hbar\nabla$
using the formula given by Trammel.\citep{Trammell_1953} Vectors
$\boldsymbol{G}\left(\boldsymbol{Q},\boldsymbol{r}\right)$ and $\boldsymbol{L}\left(\boldsymbol{Q},\boldsymbol{r}\right)$
are defined as $\boldsymbol{G}\left(\boldsymbol{Q},\boldsymbol{r}\right)=\frac{\alpha^{2}}{r_{\text{e}}}\boldsymbol{r}g\left(\boldsymbol{Q}\cdot\boldsymbol{r}\right)$
and $\boldsymbol{L}\left(\boldsymbol{Q},\boldsymbol{r}\right)=\frac{1}{2}\frac{\boldsymbol{\ell}}{\hbar}f\left(\boldsymbol{\boldsymbol{Q}}\cdot\boldsymbol{r}\right)+f\left(\boldsymbol{\boldsymbol{Q}}\cdot\boldsymbol{r}\right)\frac{1}{2}\frac{\boldsymbol{\ell}}{\hbar}$
with $g\left(x\right)=\frac{1}{ix}\left(e^{ix}-1\right)$, $f\left(x\right)=-2i\frac{d}{dx}g\left(x\right)$,
and $\boldsymbol{\ell}=-i\hbar\boldsymbol{r}\times\nabla$; $r_{\mathrm{e}}$
is the classical electron radius. We refer to the transition processes
described by the operators $\hat{f}_{\mathrm{C}}$, $\hat{f}_{\mathrm{E}}$,
$\hat{f}_{\mathrm{O}}$, and $\hat{f}_{\mathrm{S}}$ as C-, E-, O-,
and S-transition, respectively.

\begin{figure}
\begin{centering}
\includegraphics[scale=0.5]{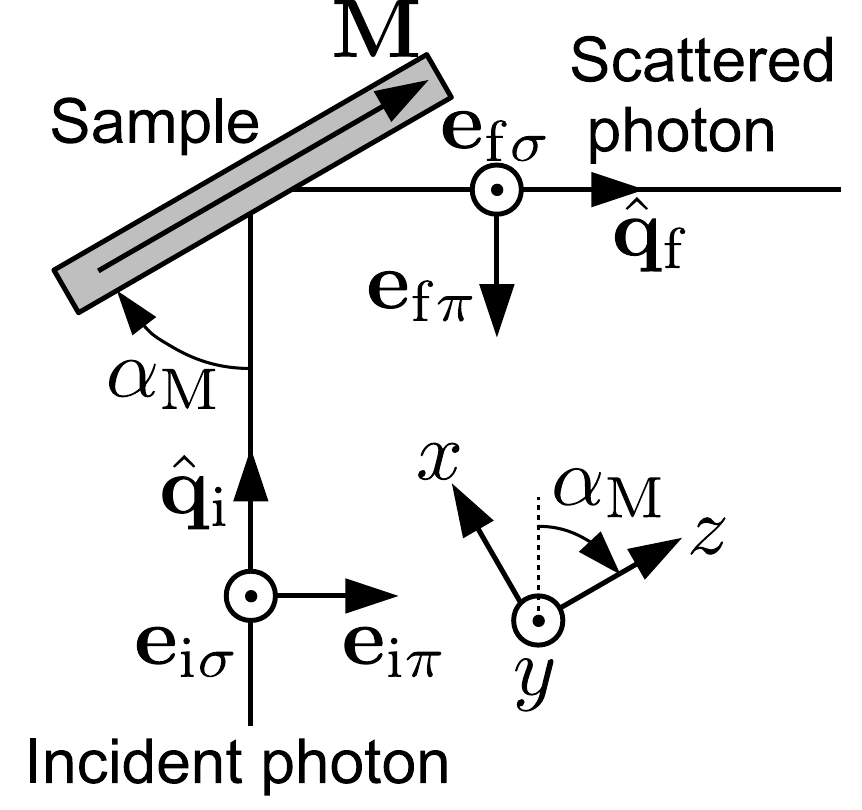}
\par\end{centering}
\caption{Schematics of scattering geometry in experiment. Coordinate and polarization
vectors are defined as shown. $\alpha_{\mathrm{M}}$ is the angle
between the magnetization vector $\boldsymbol{M}$ and the incident
propagation vector $\boldsymbol{q}_{\mathrm{i}}$. The propagation
vectors $\boldsymbol{q}_{\mathrm{i}}$ and $\boldsymbol{q}_{\mathrm{f}}$
of incident and emitted x-ray are perpendicular to each other.\label{figscattering_geometry}}
\end{figure}

The XRS-MCD experiment was carried out in the scattering geometry
shown in figure \ref{figscattering_geometry}. The wave vector $\boldsymbol{q}_{\mathrm{f}}$
is perpendicular to the incident wave vector $\boldsymbol{q}_{\mathrm{i}}$.
In the experiment, the polarization of the emitted photon is not detected
while the incident photon polarization is controlled. The polarization
of the incident photon can be characterized by Stokes parameters $P_{1}$,
$P_{2}$, and $P_{3}$.\citep{1982} Since the magnitude of the C-transition
matrix elements are much larger than the others, the total scattering
intensity $I_{\mathrm{TOT}}=I\left(P_{1},P_{2},P_{3}\right)+I\left(P_{1},-P_{2},P_{3}\right)$
is approximately given only by the C-transition as $(1+P_{3})I_{\text{CC}}^{\sigma\sigma,\sigma\sigma}$
and the MCD signal $I_{\text{MCD}}=I\left(P_{1},\left|P_{2}\right|,P_{3}\right)-I\left(P_{1},-\left|P_{2}\right|,P_{3}\right)$
is given by $2\left|P_{2}\right|{\rm Im}(I_{\text{EC}}^{\sigma\pi,\sigma\sigma}+I_{\text{OC}}^{\sigma\pi,\sigma\sigma}+I_{\text{SC}}^{\sigma\pi,\sigma\sigma})$,
where $\mathrm{Im}I$ represents the imaginary part of $I$; $I_{\text{EC}}^{\sigma\pi,\sigma\sigma}$
is given by 
\begin{equation}
I_{\text{EC}}^{\sigma\pi,\sigma\sigma}(\Delta E)=I_{0}\sum_{\Phi_{\text{f}}}\sum_{i,i^{\prime}}\delta_{E}\bigl\langle\Phi_{\text{f}}|\hat{f}_{\text{E}}^{\sigma\pi}(\boldsymbol{x}_{i^{\prime}})|\Phi_{\text{i}}\bigr\rangle^{*}\bigl\langle\Phi_{\text{f}}|\hat{f}_{\text{C}}^{\sigma\sigma}\left(\boldsymbol{x}_{i}\right)|\Phi_{\text{i}}\bigr\rangle,\label{eq:I_EC}
\end{equation}
with $I_{0}=r_{\mathrm{e}}^{2}\omega_{\mathrm{f}}/\omega_{\mathrm{i}}$.
Here, the polarization vectors in the transition operator $\hat{f}_{\text{E}}^{\sigma\pi}$
are specified as $\boldsymbol{e}_{\mathrm{f}}=\boldsymbol{e}_{\mathrm{f}\sigma}$
and $\boldsymbol{e}_{\mathrm{i}}=\boldsymbol{e}_{\mathrm{i}\pi}$.
$I_{\text{CC}}^{\sigma\sigma,\sigma\sigma}$, $I_{\text{OC}}^{\sigma\pi,\sigma\sigma}$
, and $I_{\text{SC}}^{\sigma\pi,\sigma\sigma}$ are also given in
the same manner.

The XRS intensity may be described as the sum of the scattering intensity
from each scattering site. We define the atomic transition matrix
elements as\\
 $F_{\mathrm{E}}^{\xi\eta}=\sum_{s_{z}}\int\psi_{\xi}^{*}\left(\boldsymbol{r},s_{z}\right)\hat{f}_{\mathrm{E}}^{\sigma\pi}(\boldsymbol{x})\psi_{\eta}\left(\boldsymbol{r},s_{z}\right)\mathrm{d}\boldsymbol{r}$,
where indices $\xi$ and $\eta$ refer to one of the spin-orbitals
in the 3d states and the 2p or 3p states at the scattering site, respectively;
$s_{z}$ represents the spin magnetic quantum number. $F_{\mathrm{C}}^{\xi\eta}$,
$F_{\mathrm{O}}^{\xi\eta}$, and $F_{\mathrm{S}}^{\xi\eta}$ are also
defined in the same manner. In the following, the orbital and spin
magnetic quantum number of the spin-orbital $\xi$($\eta$) are expressed
as $m_{\xi(\eta)}$ and $z_{\xi(\eta)}$. Thus, the wave functions
$\psi_{\xi}\left(\boldsymbol{r},s_{z}\right)$ might be given as a
product of the radial wave-function $R_{\xi}\left(r\right)=R_{n_{\xi}l_{\xi}}\left(r\right)$,
spherical harmonics $Y_{\xi}\left(\hat{\boldsymbol{r}}\right)=Y_{l_{\xi}m_{\xi}}\left(\hat{\boldsymbol{r}}\right)$,
and spin function $\chi_{\xi}\left(s_{z}\right)=\chi_{z_{\xi}}\left(s_{z}\right)$,\citep{Varshalovich_1988}
where $n_{\xi}=3$, $l_{\xi}=2$, $m_{\xi}=0,\pm1,\pm2$, and $z_{\xi}=\pm1/2$.
The wave function $\psi_{\eta}\left(\boldsymbol{r},s_{z}\right)$
is also written in the similar form. Functions $e^{i\boldsymbol{Q}\cdot\boldsymbol{r}}$,
$g\left(\boldsymbol{Q}\cdot\boldsymbol{r}\right)$, and $f\left(\boldsymbol{Q}\cdot\boldsymbol{r}\right)$
can be written in the spherical harmonic expansion forms $4\pi\sum_{lm}Y_{lm}^{*}\left(\hat{\boldsymbol{Q}}\right)i^{l}w_{l}\left(Qr\right)Y_{lm}\left(\hat{\boldsymbol{r}}\right)$,
where $w_{l}\left(x\right)$ is $j_{l}\left(x\right)$, $g_{l}\left(x\right)$,
and $f_{l}\left(x\right)$; $j_{l}\left(x\right)$ is the spherical
Bessel function with degree $l$, $g_{l}\left(x\right)=\frac{1}{x}\int_{0}^{x}j_{l}\left(t\right)dt$
and $f_{l}\left(x\right)=\frac{2}{x^{2}}\int_{0}^{x}tj_{l}\left(t\right)dt$,
respectively.

For the scattering geometry as shown in figure \ref{figscattering_geometry},
the atomic transition matrix elements are written as \begin{subequations}
\begin{align}
 & F_{\mathrm{C}}^{\xi\eta}=4\pi\sum_{lm}i^{l}\tilde{j}_{l}\left(Q\right)Y_{lm}^{*}\left(\hat{\boldsymbol{Q}}\right)\left(Y_{lm}\right)_{\xi\eta},\label{eq:atmC}\\
 & F_{\text{E}}^{\xi\eta}=\frac{4\pi i\Delta E}{\alpha m_{\text{e}}c^{2}}\sum_{lm\mu}i^{l}\tilde{g}_{l}\left(Q\right)Y_{lm}^{*}\left(\hat{\boldsymbol{Q}}\right)e_{\text{f}\sigma}^{\mu}\left(\hat{r}_{\mu}Y_{lm}\right)_{\xi\eta},\label{eq:atmE}\\
 & F_{\text{O}}^{\xi\eta}=-\frac{4\pi iE_{Q}}{m_{\text{e}}c^{2}}\sum_{lm\mu}i^{l}\tilde{f}_{l}\left(Q\right)Y_{lm}^{*}\left(\hat{\boldsymbol{Q}}\right)\frac{v^{\mu}\left(\left[\ell_{\mu}Y_{lm}\right]\right)_{\xi\eta}}{4\hbar},\label{eq:atmO}\\
 & F_{\text{S}}^{\xi\eta}=\frac{4\pi i\bar{E}}{m_{\text{e}}c^{2}}\sum_{lm\mu}i^{l}\tilde{j}_{l}\left(Q\right)Y_{lm}^{*}\left(\hat{\boldsymbol{Q}}\right)\hat{q}_{\text{f}}^{\mu}\left(\frac{\sigma_{\mu}}{2}Y_{lm}\right)_{\xi\eta},\label{eq:atmS}
\end{align}
\end{subequations} where $\left[\ell_{\mu}Y_{lm}\right]=\ell_{\mu}Y_{lm}+Y_{lm}\ell_{\mu}$
and the index $\mu$ runs over $1,0,-1$; $\tilde{j}_{l}\left(Q\right)$,
$\tilde{g}_{l}\left(Q\right)$, and $\tilde{f}_{l}\left(Q\right)$
are radial integrals $\int R_{\xi}\left(r\right)h\left(r\right)R_{\eta}\left(r\right)r^{2}d\boldsymbol{r}$,
where $h\left(r\right)$ is $j_{l}\left(Qr\right)$, $\frac{\alpha^{2}}{r_{\mathrm{e}}}rg_{l}\left(Qr\right)$,
and $f_{l}\left(Qr\right)$, respectively. The bracket $\left(A\right)_{\xi\eta}$
represents the directional integral\\
 $\sum_{s_{z}}\int Y_{\xi}^{*}\left(\hat{\boldsymbol{r}}\right)\chi_{\xi}\left(s_{z}\right)A\left(\hat{\boldsymbol{r}},\boldsymbol{s}\right)Y_{\eta}\left(\hat{\boldsymbol{r}}\right)\chi_{\eta}\left(s_{z}\right)d^{2}\hat{\boldsymbol{r}}$.
The vector components $a^{\mu}$($a_{\mu}$) represent the spherical
contravariant (covariant) components of vector $\boldsymbol{a}$.\citep{Varshalovich_1988}
$\boldsymbol{v}$ is given by the vector product $\boldsymbol{e}_{\text{f}\sigma}\times\hat{\boldsymbol{Q}}$.
The contravariant components $\left(a^{1},a^{0},a^{-1}\right)$ of
the vectors $\boldsymbol{e}_{\mathrm{f}\sigma}$, $\hat{\boldsymbol{q}}_{\mathrm{f}}$,
$\boldsymbol{v}$, and $\hat{\boldsymbol{Q}}$ can be written, as
$(\frac{i}{\sqrt{2}},0,\frac{i}{\sqrt{2}}$), $(\frac{C_{\alpha}}{\sqrt{2}},S_{\alpha},\frac{-C_{\alpha}}{\sqrt{2}})$
, $(-\frac{C_{\beta}}{\sqrt{2}},-S_{\beta},\frac{C_{\beta}}{\sqrt{2}}$),
and $(-\frac{S_{\beta}}{\sqrt{2}},C_{\beta},\frac{S_{\beta}}{\sqrt{2}})$,
respectively, where $C_{\alpha}=\cos\alpha_{\mathrm{M}}$, $S_{\alpha}=\sin\alpha_{\mathrm{M}}$,
$C_{\beta}=\cos(\alpha_{\mathrm{M}}+\gamma)$, and $S_{\beta}=\sin(\alpha_{\mathrm{M}}+\gamma)$
with $\gamma\approx\pi/4$.

The angle $\alpha_{\mathrm{M}}=135^{\circ}$ is found to be a special
angle like as in XMD,\citep{Blume_1988,Lovesey_1987,Laundy_1991}
which is called S-position. Because $\boldsymbol{Q}=\left(0,-1,0\right)$,
thereby $Y_{lm}\left(\hat{\boldsymbol{Q}}\right)=\sqrt{\left(2l+1\right)/4\pi}\delta_{m0}$,
the C-transition conserves both of the total orbital angular momentum
$L_{z}$ and the total spin angular momentum $S_{z}$. The E- and
O- transition change the orbital angular moment $L_{z}$ into $L_{z}\pm1$
because the vector components $e_{\text{f}\sigma}^{0}$ and $v^{0}$
are zero so that only the matrix elements $\left(\hat{r}_{\pm1}Y_{l0}\right)_{\xi\eta}$
and $\left(\left[\ell_{\pm1}Y_{l0}\right]\right)_{\xi\eta}$ could
be non-zero, while they conserve $S_{z}$. The S-transition conserves
$L_{z}$, but changes $S_{z}$ into any of $S_{z}\pm1$ and $S_{z}$.
Inevitably, if we can assume that the electron system conserves the
$z$ component of the total angular momentum $J_{z}=L_{z}+S_{z}$
around the scattering site when the electron system is not affected
by the external perturbations, the interference terms $I_{\text{EC}}^{\sigma\pi,\sigma\sigma}$
and $I_{\text{OC}}^{\sigma\pi,\sigma\sigma}$ would be zero. The terms
which involve the spin-off-diagonal S-transition in $I_{\mathrm{SC}}^{\sigma\pi,\sigma\sigma}$
also would be zero. Even in case the angular momentum $J_{z}$ is
not conserved, if the powder approximation are allowed, the effect
of such interference terms would not appear on the scattering intensity.
Consequently, only the terms which involves the spin-diagonal S-transition
in $I_{\text{SC}}^{\sigma\pi,\sigma\sigma}$ can contribute to the
XRS intensity as MCD signal. Further, the term $I_{\text{S\ensuremath{\uparrow}C\ensuremath{\downarrow}}}^{\sigma\pi,\sigma\sigma}$
in $I_{\text{SC}}^{\sigma\pi,\sigma\sigma}$ consisting of the S$\uparrow$-transition,
in which an up-spin electron is excited, and the C$\downarrow$-transition,
in which a down electron is excited, and the term $I_{\text{S\ensuremath{\downarrow}C\ensuremath{\uparrow}}}^{\sigma\pi,\sigma\sigma}$
consisting of the S$\downarrow$-transition and the C$\uparrow$-transition
would not contribute to the XRS intensity, because they have the same
magnitude but have the opposite sign to each other. Therefore, putting
$M_{m}=\sum_{\ell}M_{m}^{\left(\ell\right)}=\sum_{\ell}i^{\ell}\sqrt{4\pi\left(2l+1\right)}\tilde{j}_{\ell}\left(Q\right)\left(Y_{\ell0}\right)_{2m,1m},$
the term $I_{\text{SC}}^{\sigma\pi,\sigma\sigma}\left(\Delta E\right)$
can be simplified as
\begin{align}
 & I_{\text{SC}}^{\sigma\pi,\sigma\sigma}\left(\Delta E\right)=\frac{iI_{0}}{\sqrt{2}}\frac{\bar{E}}{m_{\text{e}}c^{2}}\sum_{\Phi_{\text{f}}}\delta\left(\Delta E+E_{\text{i}}-E_{\text{f}}\right)\nonumber \\
 & \times\sum_{mm^{\prime}}M_{m^{\prime}}^{*}M_{m}\left\langle \Phi_{\text{i}}\left|p_{m^{\prime}\uparrow}^{\dagger}d_{m^{\prime}\uparrow}\right|\Phi_{\text{f}}\right\rangle \left\langle \Phi_{\text{f}}\left|d_{m\uparrow}^{\dagger}p_{m\uparrow}\right|\Phi_{\text{i}}\right\rangle \nonumber \\
 & -\left\{ \uparrow\longleftrightarrow\downarrow\right\} .\label{eq:simplified_ISC}
\end{align}
If we can assume that the 3p or 2p core states are completely occupied
in the initial state $\Phi_{\mathrm{i}}$, the integrated $I_{\text{CC}}^{\sigma\sigma,\sigma\sigma}$
and $I_{\text{SC}}^{\sigma\pi,\sigma\sigma}$in terms of the transferred
energy $\text{\ensuremath{\Delta E}}$ can be related to the hole
number as\begin{subequations}
\begin{align}
 & \int_{E_{\mathrm{E}}}^{E_{\mathrm{C}}}\mathrm{d}xI_{\text{CC}}^{\sigma\sigma,\sigma\sigma}\left(x\right)=I_{0}N_{1},\label{eq:Int_ICC}
\end{align}
\begin{align}
 & \int_{E_{\mathrm{E}}}^{E_{\mathrm{C}}}\mathrm{d}x\mathrm{Im}I_{\text{SC}}^{\sigma\pi,\sigma\sigma}\left(x\right)=\frac{I_{0}}{2\sqrt{2}}\frac{\bar{E}}{m_{\text{e}}c^{2}}S_{1},\label{eq:Int_ISC}
\end{align}
\end{subequations}where $E_{\mathrm{E}}$ and $E_{\mathrm{C}}$ indicate
the transferred energy at the edge and an appropriate cutoff energy,
respectively. $N_{1}$ and $S_{1}$ are defined as $N_{1}=\sum_{m=-1}^{1}\left|M_{m}\right|^{2}\left(h_{m\uparrow}+h_{m\downarrow}\right)$
and $S_{1}=\sum_{m=-1}^{1}\left|M_{m}\right|^{2}\left(h_{m\uparrow}-h_{m\downarrow}\right)$,
and $h_{m\uparrow(\downarrow)}$ is the up (down) spin hole number
in the 3d state specified by the orbital magnetic quantum number $m$.
Thereby, we obtain
\begin{equation}
\frac{S_{1}}{N_{1}}=C\frac{\int_{E_{\mathrm{E}}}^{E_{\mathrm{C}}}\mathrm{d}xI_{\mathrm{MCD}}\left(x\right)}{\int_{E_{\mathrm{E}}}^{E_{\mathrm{C}}}\mathrm{d}xI_{\mathrm{TOT}}\left(x\right)},\label{eq:S/N}
\end{equation}
 with $C=\sqrt{2}\left(1+P_{3}\right)m_{\mathrm{e}}c^{2}/\left|P_{2}\right|\bar{E}$.

On the other hand, when the angle $\alpha_{\mathrm{M}}=0$ or $45\text{ degrees}$,
which correspond to the L or L+S position in the XMD, the XRS-MCD
signals may show complex behavior because both of $I_{\mathrm{EC}}$
and $I_{\mathrm{OC}}$ take part in the MCD signals. For the transition
metal M-edge excitation, although the transferred energy $\Delta E$
is much smaller than that for the L-edge excitation, the E-transition
is not negligible as shown later. If the transferred energy $\Delta E$
is so large that the contribution $I_{\mathrm{OC}}$ and $I_{\mathrm{SC}}$
is negligible, the sum-rules similar to those in the XAS-MCD might
be established.

\section{Model Hamiltonian and Calculation Method}

The configuration interaction (CI) calculation on the Anderson impurity
model has been applied to analyze the signals from several core-level
spectroscopic experiment on ferromagnetic nickel and have given consistent
explanations to the different spectra based on the calculated electronic
structure. \citep{Jo_1991,Tanaka_1992} Although the validity to apply
this model for discussion on the spectroscopic properties of more
strongly itinerant electron systems is not guaranteed, we exploit
the CI calculation on this model as a makeshift to demonstrate the
usefulness of the XRS-MCD in this study, because the electron-hole
interaction is so large that independent particle approximation may
not be suitable for describing the $\mathrm{M}_{2,3}$-edge excitation.

The 3d electron number of the Fe ion could be strongly fluctuating.
We assume that the 3d electrons go back and forth between the 3d states
under consideration and the electron reservoir states, which are supposed
to have d-symmetric states consisting of the 3d and/or 4s states around
the scattering site. We prepare the ten different levels $\nu_{\alpha m}$
($\alpha=0,1$ and $m=-2,-1,\ldots,2$) as electron reservoir states.
The initial electronic state might be symbolically expressed as $\left|\Phi_{\text{i}}\right\rangle =A\left|d^{8}\nu^{n_{0}+2}\right\rangle +B\left|d^{7}\nu^{n_{0}+3}\right\rangle +C\left|d^{6}\nu^{n_{0}+4}\right\rangle ,$where
$d^{m}\nu^{n}$ represents the configuration, in which $m$ electrons
occupy the 3d states under consideration and $n$ electrons do the
reservoir states. $A\left|d^{m}\nu^{n}\right\rangle $ represents
the linear combination $\sum_{c}A_{c}\left|d^{m}\nu^{n}\right\rangle _{c}$
over the configurations belonging to the states specified as $d^{m}\nu^{n}$.\footnote{We assumed $n_{0}=4$, which gives most plausible results.}

We assume the model Hamiltonian for simulating the electronic state
as
\begin{align}
H & =\sum_{\xi}E_{d}n_{d\xi}+\sum_{\alpha\xi}E_{\alpha}n_{\alpha\xi}+\sum_{\alpha\xi}V\left(d_{\xi}^{\dagger}c_{\alpha\xi}+c_{\alpha\xi}^{\dagger}d_{\xi}\right)\nonumber \\
 & +U_{dd}\sum_{\xi<\xi^{\prime}}n_{d\xi}n_{d\xi^{\prime}}+U_{pd}\sum_{\xi\eta}n_{d\xi}n_{p\eta}\nonumber \\
 & +H_{dd}\left(F_{dd}^{2},F_{dd}^{4}\right)+H_{pd}\left(F_{pd}^{2},G_{pd}^{1},G_{pd}^{3}\right)\nonumber \\
 & -\Delta_{\text{mol}}z_{\xi}n_{d\xi}+H_{d\text{SO}}\left(\zeta_{d}\right)+H_{p\text{SO}}\left(\zeta_{p}\right),\label{eq:H}
\end{align}
where, $d_{\xi}^{\dagger}$, $d_{\xi}$, and $n_{d\xi}$ represent
the creation, annihilation, and number operators for the spin-orbital
$\xi$ in the 3d state at the site under consideration. $c_{\alpha\xi}^{\dagger}$,
$c_{\alpha\xi}$, and $n_{\alpha\xi}$ represent the creation, annihilation,
and number operators for the spin orbital $\nu_{\alpha m_{\xi}z_{\xi}}$
in the reservoir states. $n_{p\eta}$ represents the number operator
for the spin-orbital $\eta$ in the 2p or 3p states. The parameters
$E_{d}$ and $E_{\alpha}$ representing the one electron level are
assumed to be $0$ eV, $-0.2\times\alpha$ eV. The parameters $U_{dd}$
and $U_{pd}$ corresponding the averaged 3d-3d and 2p(3p)-3d Coulomb
interaction are assumed to be $3.5$eV and $5.0$ ($3.5$) eV. The
hybridization $V$ is assumed to be $1.1$ eV. The Slater integrals
$F_{dd}^{2}$, $F_{dd}^{4}$, $F_{pd}^{2}$, $G_{pd}^{1}$, and $G_{pd}^{3}$
are assumed to be 80\% of the atomic values. The parameters $\zeta_{d}$
and $\zeta_{p}$ of the spin-orbit coupling $H_{d\mathrm{SO}}$and
$H_{p\mathrm{SO}}$ are assumed to be the atomic values. These atomic
values are calculated by using Cowan code.\citep{cowan1981theory}
We add the molecular field term $-\Delta_{\text{mol}}z_{\xi}n_{d\xi}$
in order to simulate the ferromagnetic ground state. The parameter
$\Delta_{\mathrm{mol}}$ is assumed to be $1.9$eV, which corresponds
to the observed exchange splitting value $\varepsilon\left(\mathrm{H}_{25\uparrow}\right)-\varepsilon\left(\mathrm{H}_{25\downarrow}\right)$.\citep{Santoni_1991}
We numerically diagonalize the Hamiltonian to obtain the initial ground
state, in which the d electron number, spin moment, and orbital moment
are about $7.0$, $2.2\mu_{\mathrm{B}}$, and $0.054\mathrm{\mu_{B}}$,
respectively. The weights $A$, $B$, and $C$ are $\left|A\right|^{2}=23.7\%$,
$\left|B\right|^{2}=55.7\%$, and $\left|C\right|^{2}=20.6\%$, which
may be consistent with the stronger itinerancy than ferromagnetic
Ni.\citep{Jo_1991,Tanaka_1992}

Scattering operators $\hat{f}_{\mathrm{E}}^{\sigma\pi}$ can be expressed
in the second quantization form using the atomic transition matrix
elements $F_{\mathrm{E}}^{\xi\eta}$: $\hat{f}_{\mathrm{E}}^{\sigma\pi}=\sum_{\xi\eta}F_{\mathrm{E}}^{\xi\eta}d_{\xi}^{\dagger}p_{\eta}$
within the model used. The terms $I_{\mathrm{EC}}^{\sigma\pi,\sigma\sigma}$
are given by $I_{\text{EC}}^{\sigma\pi,\sigma\sigma}=\frac{I_{0}}{2\pi i}\left\langle \Phi_{\text{i}}\left|\hat{f}_{\text{E}}^{\dagger}\left[R\left(z\right)-R\left(z^{*}\right)\right]\hat{f}_{\text{C}}\right|\Phi_{\text{i}}\right\rangle ,$where
$R\left(z\right)=1/\left(z-H\right)$ with $z=\Delta E+E_{\text{i}}-H-i\Gamma$
and $z^{*}=\Delta E+E_{\text{i}}-H+i\Gamma$. The other terms $I_{\mathrm{CC}}^{\sigma\sigma,\sigma\sigma}$,
$I_{\mathrm{OC}}^{\sigma\pi,\sigma\sigma}$, and $I_{\mathrm{SC}}^{\sigma\pi,\sigma\sigma}$
also can be given in the same manner. To calculate these terms we
can use the recursion method with assuming that the final states be
described as $\left|\Phi_{\text{f}}\right\rangle =A^{\prime}\left|p^{5}d^{9}\nu^{n_{0}+2}\right\rangle +B^{\prime}\left|p^{5}d^{8}\nu^{n_{0}+3}\right\rangle +C^{\prime}\left|p^{5}d^{7}\nu^{n_{0}+4}\right\rangle $,
where $p^{n}$ indicates the states that $n$ electrons are accommodated
in the 2p or 3p state. It is well known that the term-dependent core-hole
lifetime due to the 3p-3d3d super-Coster-Kroning decay plays significant
roles for explaining the observed spectral shape in the M-edge spectroscopy.\citep{Okada_1993}
Such core-hole decay processes are not taken into account in our model
Hamiltonian. Taguchi et al. assumed that the core-hole lifetime broadening
$\Gamma$ of 3p hole is linear on the relative excitation energy in
order to investigate the emission spectra from manganese oxides.\citep{Taguchi_1997}
Although we have no substantial reasons, we assume the broadening
$\Gamma$ linearly depending on the relative excitation energy, when
comparing the calculated and observed spectra at the $M_{2,3}$ edge.

As shown later, we obtain plausible results for both of the $L_{2,3}$
and $M_{2,3}$ edges XRS spectra. The spectral shape is not sensitive
on the model parameters as far as we use the initial state in which
the 3d electron number is about $7.0$ and the spin moment is about
$2.2\mathrm{\mu_{B}}$. However, the validity of the calculated spectra
based on the above mentioned approximation is probably quite limited.
In the spectral shape at the $M_{2,3}$ edge, several inconsistencies
are found between the observation and the calculation. Nevertheless,
we hope that the results are of value to provide insight into the
XRS-MCD and understand its usefulness.

\section{Results and Discussions}

\subsection{Fe $L_{2,3}$ edge}

\begin{figure}
\begin{centering}
\includegraphics[clip,scale=0.75]{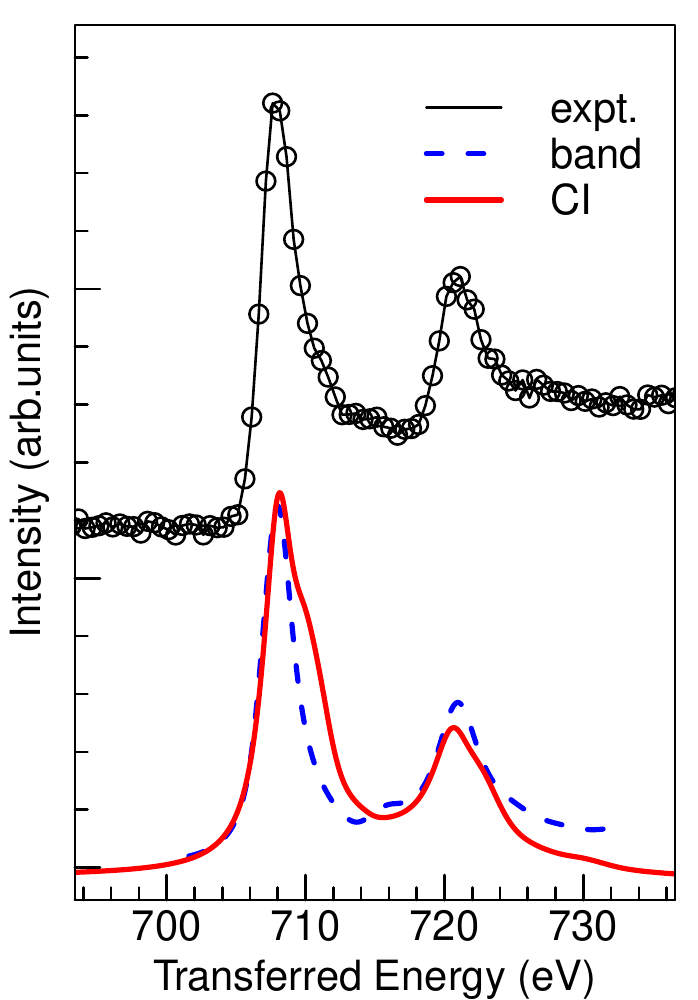}
\par\end{centering}
\caption{(color on line) Total XRS intensity as a function of the transferred
energy and the calculated intensity $I_{\mathrm{CC}}$. Thin sold
line with circle symbol represents the observed intensity without
background subtraction. Thick solid, and broken lines are the spectral
curves calculated by the CI calculation and the band structure calculation,
respectively. Lifetime broadening is assumed to be $\Gamma_{3}=1.4$
and $\Gamma_{2}=0.9$ eV. \label{fig:L_C_2theta=00003D90_a=00003D0}}
\end{figure}

In the previous papers,\citep{Hiraoka_2015,Takahashi_2015} we investigated
the XRS-MCD at the Fe $L_{2,3}$ edge within independent particle
approximation using band structure calculation based on the local
spin density approximation. At the $L_{2,3}$ edge, the dipole transitions
dominates the scattering intensity and the MCD signal, in which the
form factors $\tilde{j}_{1}\left(Q\right)$, $\tilde{g}_{0}\left(Q\right)$,
and $\tilde{f}_{1}\left(Q\right)$ are relevant. In figure \ref{fig:L_C_2theta=00003D90_a=00003D0},
we compare the total intensities calculated by the CI calculation
and the band calculation with the experimental observation.\footnote{In the experiments at the L- and M-edges , the incident photon energy
is scanned over a specific range to detect emitted photons with an
energy of 9888 eV.} Both of the calculations well reproduce the observed spectral curve.
The observed $\mathrm{L}_{3}$ peak, concentrating around the transferred
energy $705\sim714\mathrm{eV}$, looks consisting of a main peak about
$708\mathrm{eV}$ and a shoulder structure around $712\mathrm{eV}$.
This shoulder structure seems not to be properly reproduced by the
calculations: The one-body calculation does not give the shoulder
structure, on the other hand the CI calculation seems to provide too
strong intensity for the shoulder structure.

\begin{figure*}
\begin{centering}
\includegraphics[clip,scale=0.4]{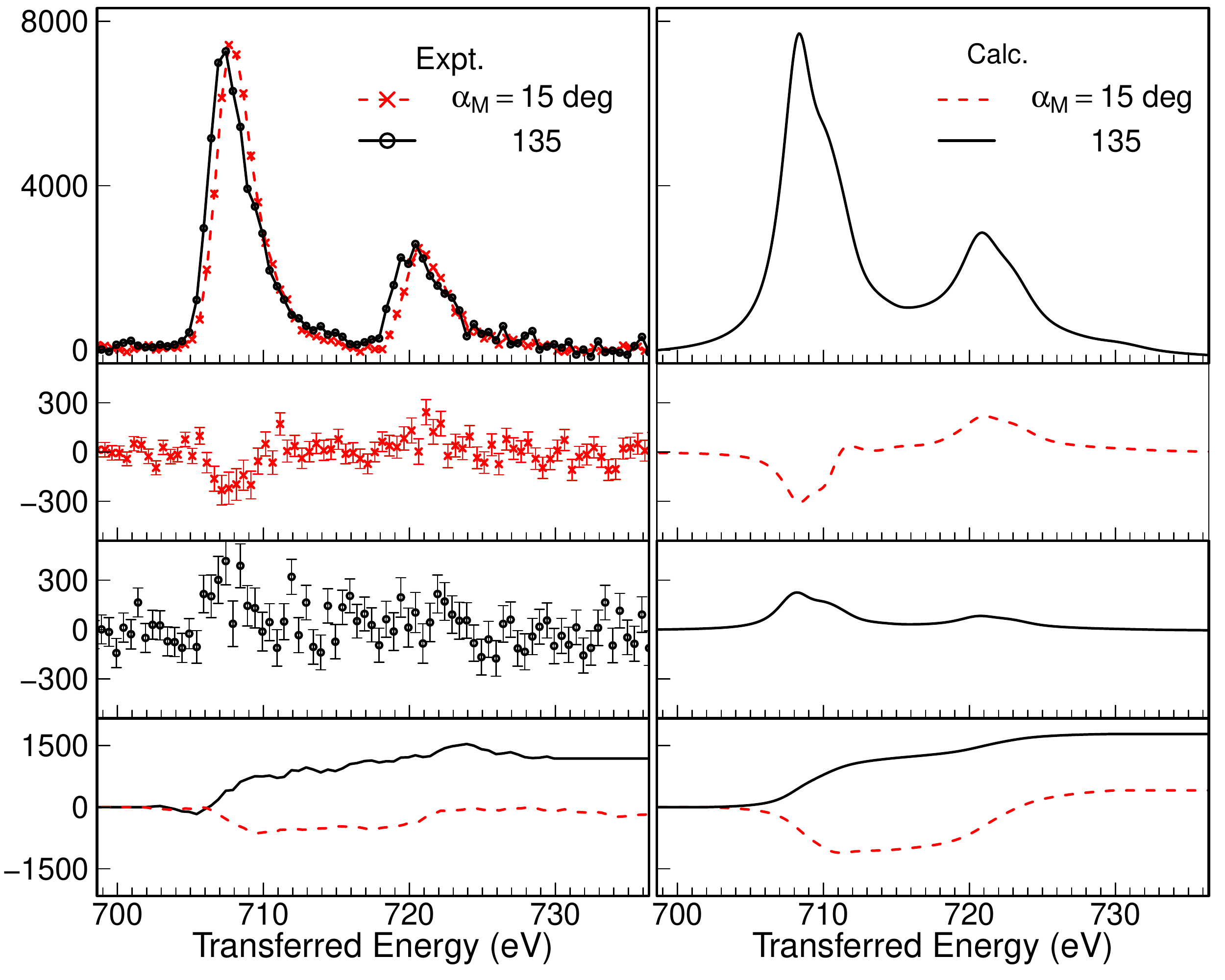}\includegraphics[clip,scale=0.4]{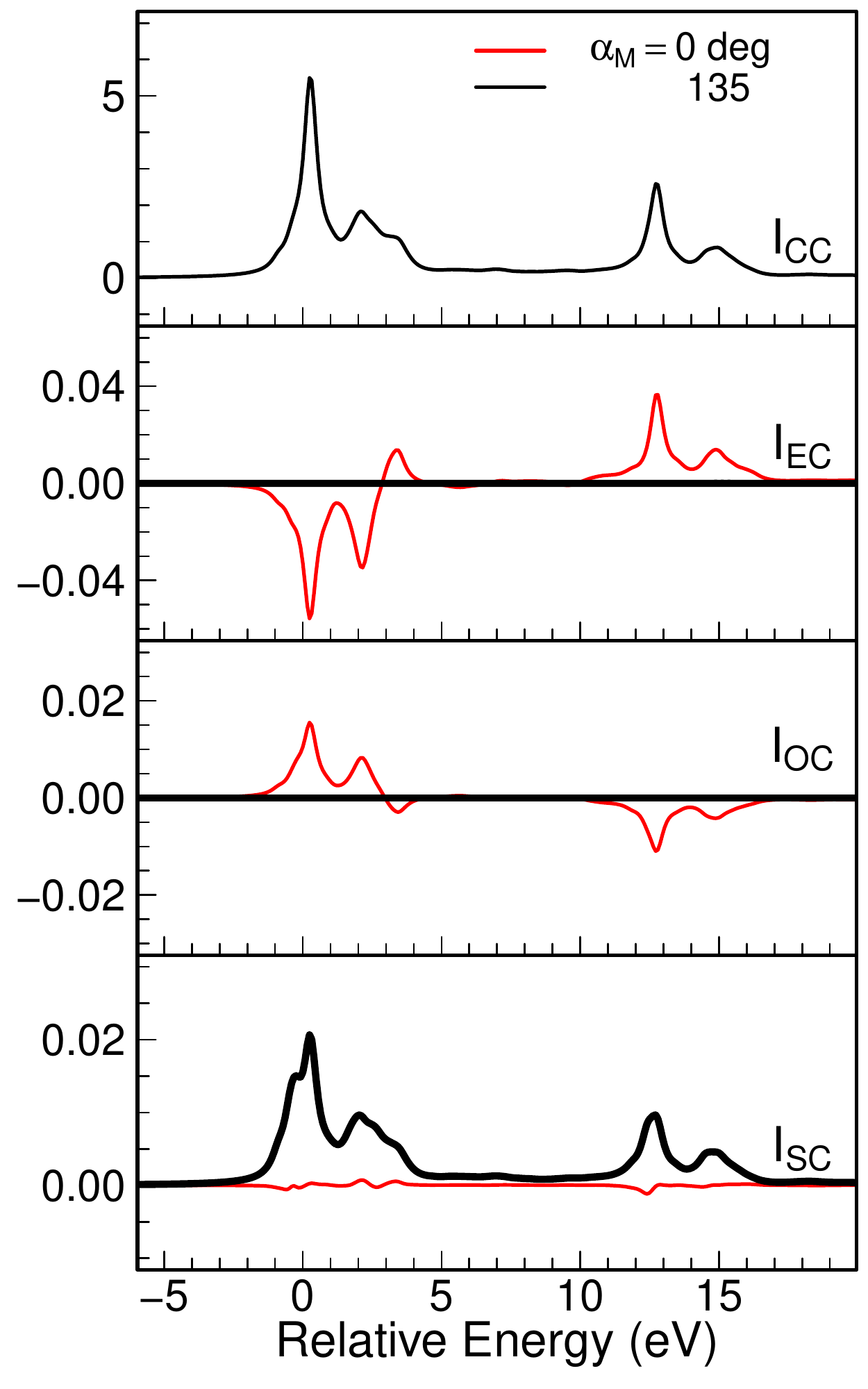}
\par\end{centering}
\caption{(color online) The observed total intensity after background subtraction
(left) and the calculated total intensity (center) are shown in the
top panels, respectively. The total XRS signals at $\alpha_{\mathrm{M}}=15^{\circ}$
and $135^{\circ}$ are shown with red dashed and black solid curves.
MCD signals at the corresponding angle $\alpha_{\mathrm{M}}$ are
shown in the second and third rows. The integrated MCD signals are
shown in bottom panels. In the right panels, the spectral curves of
the intensity $I_{\mathrm{CC}}$, $\mathrm{Im}I_{\mathrm{EC}}$, $\mathrm{Im}I_{\mathrm{OC}}$,
and $\mathrm{Im}I_{\mathrm{SC}}$ at the angles $\alpha_{\mathrm{M}}=0^{\circ}$
and $135^{\circ}$ are shown with lifetime broadening $\Gamma=0.14\mathrm{eV}$.
\label{fig:FeL_MCD-2}}
\end{figure*}
The total intensity after background subtraction and the MCD signals
are shown in figure \ref{fig:FeL_MCD-2} in comparison with those
obtained by the CI calculation. The Stokes parameters of the incident
beam polarization are assumed as $\left|P_{2}\right|=0.6$ and $P_{3}=-0.8$.
The CI calculation well reproduces the observed MCD signals both on
the relative intensity to the total intensity, the sign of MCD signal,
and their dependence on the angle $\alpha_{\mathrm{M}}$. In the
most right panels, the spectral curves of the intensity $I_{\mathrm{CC}}$,
and the MCD components $\mathrm{Im}I_{\mathrm{EC}}$, $\mathrm{Im}I_{\mathrm{OC}}$,
and $\mathrm{Im}I_{\mathrm{SC}}$ at the angle $\alpha_{\mathrm{M}}=0^{\circ}$
and $135^{\circ}$ are also shown. At the angle $\alpha_{\mathrm{M}}=0^{\circ}$
, $\mathrm{Im}I_{\mathrm{EC}}$ dominantly contributes to the total
MCD signal, while at the angle $\alpha_{\mathrm{M}}=135^{\circ}$
, $\mathrm{Im}I_{\mathrm{EC}}$ and $\mathrm{Im}I_{\mathrm{OC}}$
are completely suppressed and only $\mathrm{Im}I_{\mathrm{SC}}$ contributes
to the total MCD signal. We note that at the angle $\alpha_{\mathrm{M}}=0^{\circ}$,
the MCD component $\mathrm{Im}I_{\mathrm{SC}}$ is not completely
zero; it would be zero if we assume the powder approximation. The
results obtained by the CI calculation are essentially the same with
those calculated by the band calculation\citep{Takahashi_2015}.

\subsection{Fe $M_{2,3}$ edge}

\begin{figure}
\begin{centering}
\includegraphics[scale=0.75]{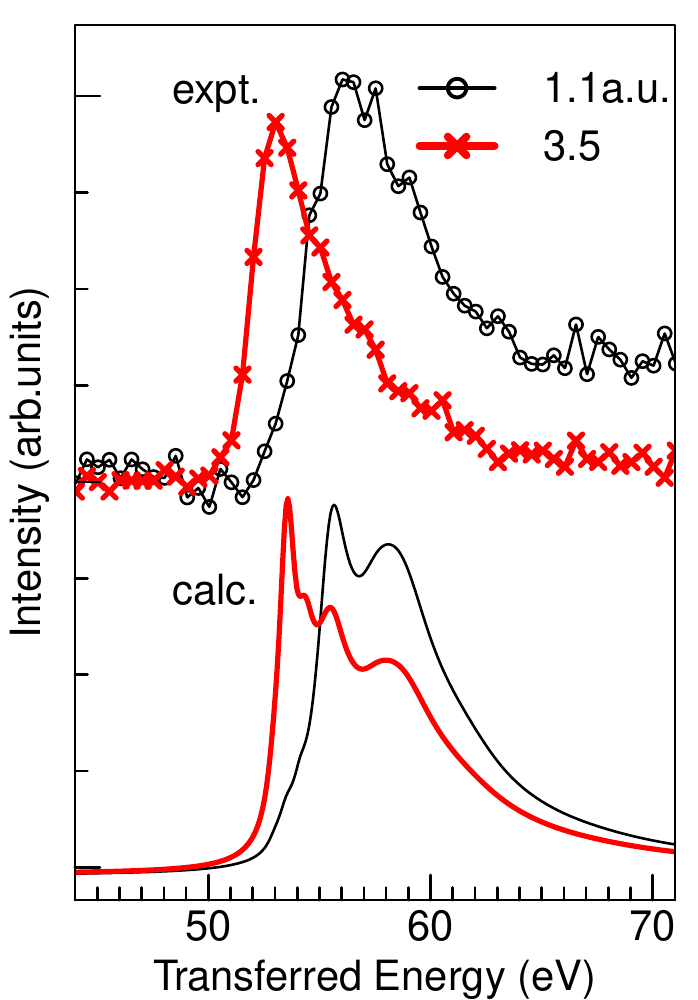}
\par\end{centering}
\caption{(Color online) Observed and calculated total XRS intensities as a
function of the transferred energy. Black thin and red thick lines
are the intensities at the scattering vector $Q=1.1$ and 3.5 a.u.
corresponding to the scattering angles $2\theta=23^{\circ}$ and $90^{\circ}$,
respectively.\label{fig:M_C_2theta=00003D23,90_a=00003D0}}
\end{figure}

The relative magnitude of $\mathrm{Im}I_{\mathrm{SC}}$ to $I_{\mathrm{CC}}$
is independent of the edges, because $I_{\mathrm{SC}}$ is directly
proportional to the 3d spin moment in accordance with eq. (\ref{eq:simplified_ISC}).
At the Fe $M_{2,3}$ edge, the MCD signal at the angle $\alpha_{\mathrm{M}}=135^{\circ}$
might be observed with the same relative magnitude to the total intensity
as at the $L_{2,3}$ edge. At the Fe $M_{2,3}$ edge the octupole
transition becomes significant as well as the dipole transition for
the high $Q$ scattering. Figure \ref{fig:M_C_2theta=00003D23,90_a=00003D0}
shows the observed and calculated total intensities as a function
of the transferred energy at the scattering angles $2\theta=23^{\circ}$
and $90^{\circ}$ , which correspond to the scattering vector $Q=1.1$
and $3.5$ a.u., respectively. At the scattering angle $2\theta=23^{\circ}$
, the intensity is dominated by the dipole transition. As the scattering
vector becomes larger, the octupole transition become dominant and
the dipole transition becomes subordinate; the intensity around the
transferred energy $53\sim55$ eV become intense and the intensity
above $55$ eV becomes weak. A similar tendency can be seen in the
XRS spectra on iron oxides.\citep{Nyrow_2014_2} In order to compare
the observation and the calculation, we naively assume that the life-time
broadening is $\Gamma=\max\left(0.1E+0.14,\,0.14\right)$ eV, where
$E$ is the relative transferred energy from the edge. Although the
calculated spectra resemble the observed one, they show discernible
inconsistencies at the scattering angle $2\theta=90^{\circ}$. In
comparison with the observed spectra, the calculated intensity above
the transferred energy $55$ eV, which is mainly caused by the dipole
transition, looks to be quite overestimated, or the intensity around
the transferred energy $53$ eV, which is mainly caused by the octupole
transition, looks to be underestimated. In order to improve the calculated
spectral curve, it might be necessary to explicitly take account of
the super-Coster-Kroning decay process into the calculation. In the
vicinity of the edge, the low-lying electron-hole-pair excitations
might be essential for the shape of the peak.\citep{Doniach_1971,Nozi_res_1974}
In spite of the noticeable deviation between the experimental observation
and the calculation, we expect that the results could give us better
understanding of the XRS-MCD.

\begin{figure*}
\begin{centering}
\includegraphics[clip,scale=0.4]{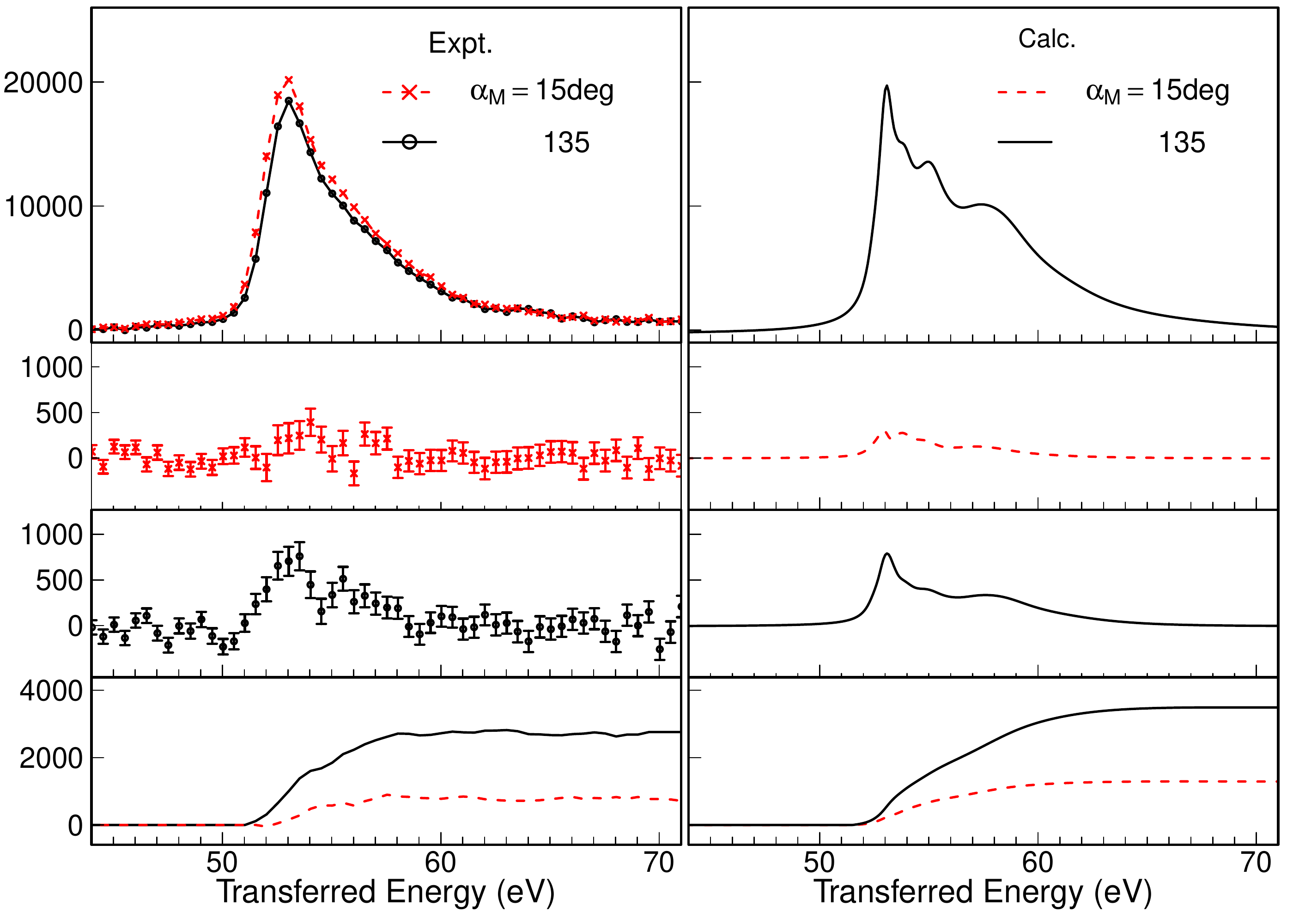}
\par\end{centering}
\caption{(color online) Observed spectra (left panel) and calculated spectra
(right panel) are shown as a function of the transferred energy. The
total intensity after subtracting the background, the MCD signal,
and the integrated MCD signal are shown in top, middle, and bottom
panels. The signals for the angle $\alpha_{\mathrm{M}}=15^{\circ}$
and $135^{\circ}$ are indicated by red cross symbols and broken lines,
and the black circle symbols and solid lines. \label{fig:FeM_MCD}}
\end{figure*}

Left panel in figure \ref{fig:FeM_MCD} shows the observed spectra
as a function of the transferred energy at the angles $\alpha_{\mathrm{M}}=15^{\circ}$
and $135^{\circ}$. Contrasting to the L-edge spectra, the spectral
curve of the MCD signal at $\alpha_{\mathrm{M}}=15^{\circ}$ is rather
simple: its magnitude is very weak and the shape is similar to that
at $\alpha_{\mathrm{M}}=135^{\circ}$, which are also similar to the
total intensity. This might suggest that the contribution of $\mathrm{Im}I_{\mathrm{EC}}$
and $\mathrm{Im}I_{\mathrm{OC}}$ are suppressed and $\mathrm{Im}I_{\mathrm{SC}}$
dominates the MCD signal. Right panel shows the calculated spectra
corresponding to the observation with the polarization parameters
$\left|P_{2}\right|=0.6$ and $P_{3}=-0.8$. The sign of the MCD signal,
the relative magnitude of the MCD signal to the total intensity and
the shape of the spectral curves are rather well reproduced by the
calculation.
\begin{figure*}
\begin{centering}
\includegraphics[clip,scale=0.4]{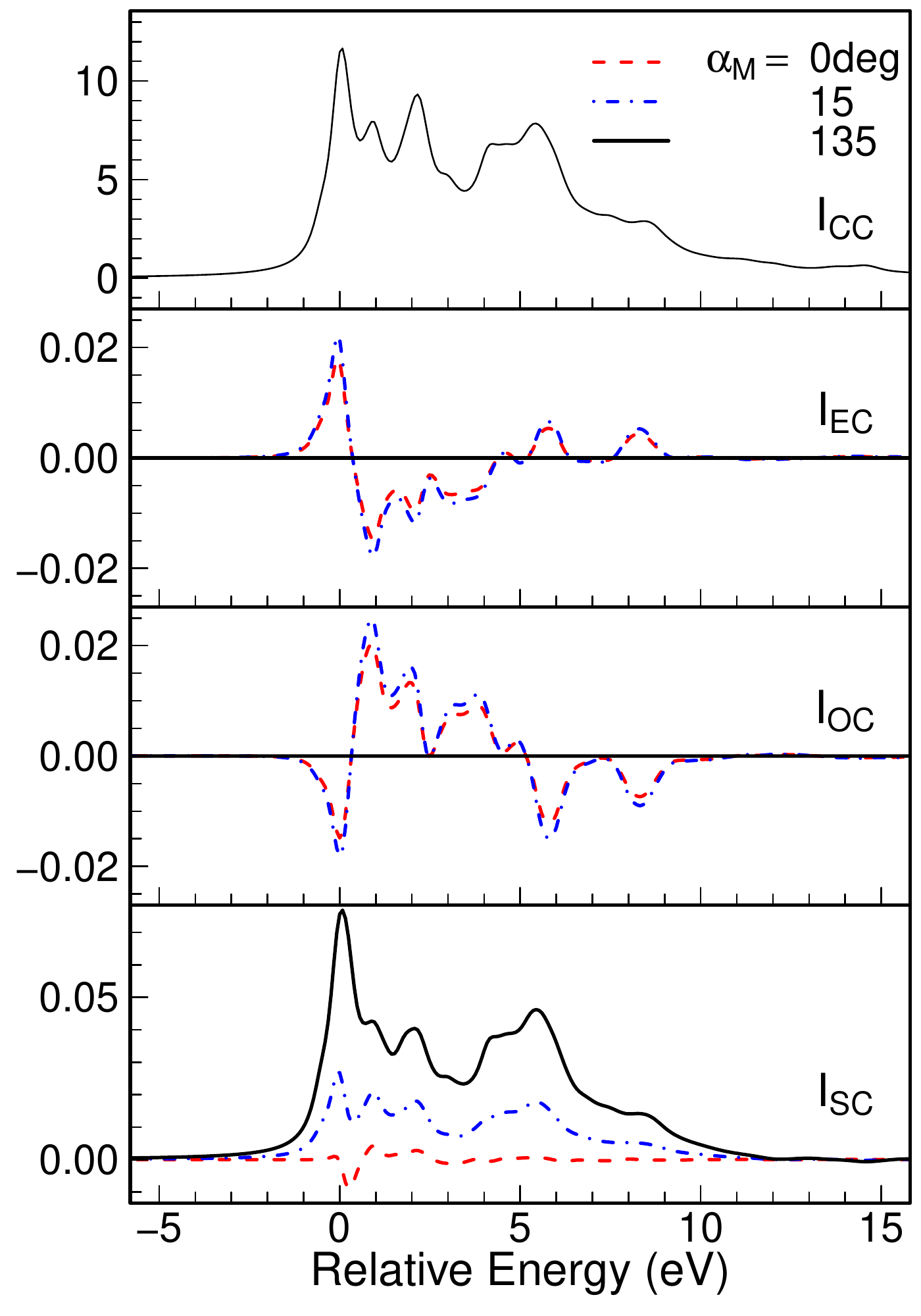}\includegraphics[clip,scale=0.4]{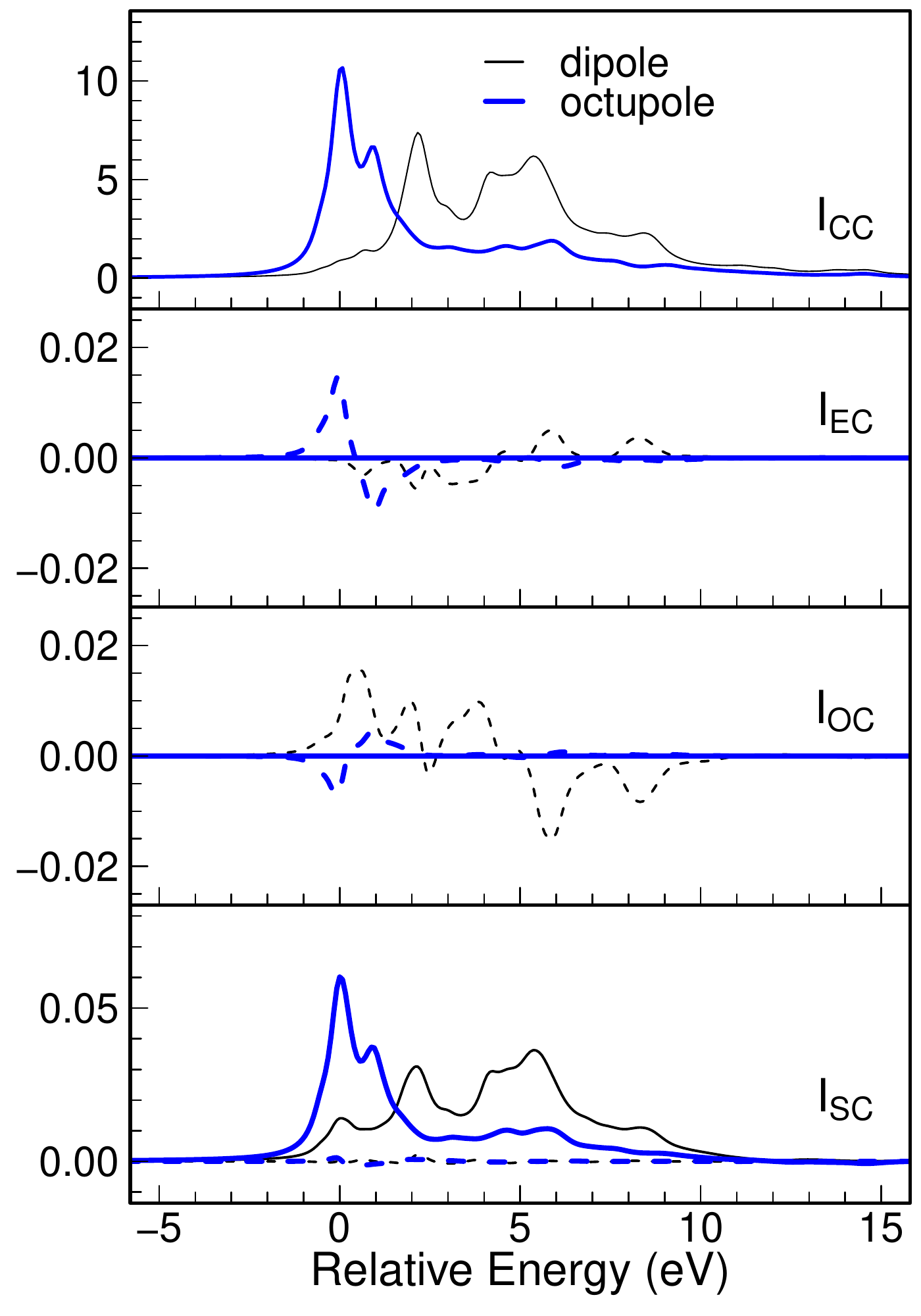}
\par\end{centering}
\caption{(Color online) Left panels show the intensities $I_{\mathrm{CC}}$,
$\mathrm{Im}I_{\mathrm{EC}}$, $\mathrm{Im}I_{\mathrm{OC}}$, and
$\mathrm{Im}I_{\mathrm{SC}}$ from top to bottom as a function of
relative energy of the final states at $\alpha_{\mathrm{M}}=0$ (dashed
line), $15^{\circ}$ (dot-dash line) and $135^{\circ}$ (solid line).
Right panels show the decomposed intensities into the parts due to
the dipole (thin line) and octupole (thick line) transition processes.
Dashed and solid curves indicate the results at the angle $\alpha_{\mathrm{M}}=0^{\circ}$
and $135^{\circ}$. The life time broadening is assumed to be $\Gamma=0.14$
eV.\label{fig:FeM_MCD_calc}}
\end{figure*}

The left panel in figure \ref{fig:FeM_MCD_calc} shows the intensities
$I_{\mathrm{CC}}$ $\mathrm{Im}I_{\mathrm{EC}}$, $\mathrm{Im}I_{\mathrm{OC}}$,
and $\mathrm{Im}I_{\mathrm{SC}}$ as a function of the relative energy
of the final states at the angle $\alpha_{\mathrm{M}}=0^{\circ}$,
$15^{\circ}$, and $135^{\circ}$. It is found that $\mathrm{Im}I_{\mathrm{EC}}$
and $\mathrm{Im}I_{\mathrm{OC}}$ can be almost canceled out to each
other near $\alpha_{\mathrm{M}}=0^{\circ}$. Consequently, the the
MCD signal at $\alpha_{\mathrm{M}}=15^{\circ}$ are dominated by $\mathrm{Im}I_{\mathrm{SC}}$.
At the L-edge, this cancellation is insufficient: $\mathrm{Im}I_{\mathrm{EC}}$
dominates the MCD signals at $\alpha_{\mathrm{M}}=15^{\circ}$. At
$\alpha_{\mathrm{M}}=135^{\circ}$, the MCD signals due to $\mathrm{Im}I_{\mathrm{EC}}$
and $\mathrm{Im}I_{\mathrm{OC}}$ are completely suppressed, so $\mathrm{Im}I_{\mathrm{SC}}$
alone contributes to the MCD signals. Thus, the MCD signals reflect
only the spin polarization in the 3d orbitals of the orbital magnetic
quantum number $m=0,\pm1$. It may be worth noting again that $\mathrm{Im}I_{\mathrm{SC}}$
is not identically zero even at the angle $\alpha_{\mathrm{M}}=0^{\circ}$
.

The left panel in figure \ref{fig:FeM_MCD_calc} shows the calculated
intensities due to the dipole transition alone and the octupole transition
alone. The dipole and octupole transitions dominate the intensities
in the range of $2.5\sim10$ eV and the range of $0\sim2.5$ eV, respectively.
The effect of the interference between them does not look significant
in the intensity $I_{\mathrm{CC}}$. On the other hand it could not
be ignored for producing the spectral structure of the MCD signal.
Therefore, the detailed information about the electronic structure
might be obtained from the analysis of the XRS-MCD signal.

\subsection{$M_{1}$ edge}

\begin{figure}
\begin{centering}
\includegraphics[clip,scale=0.4]{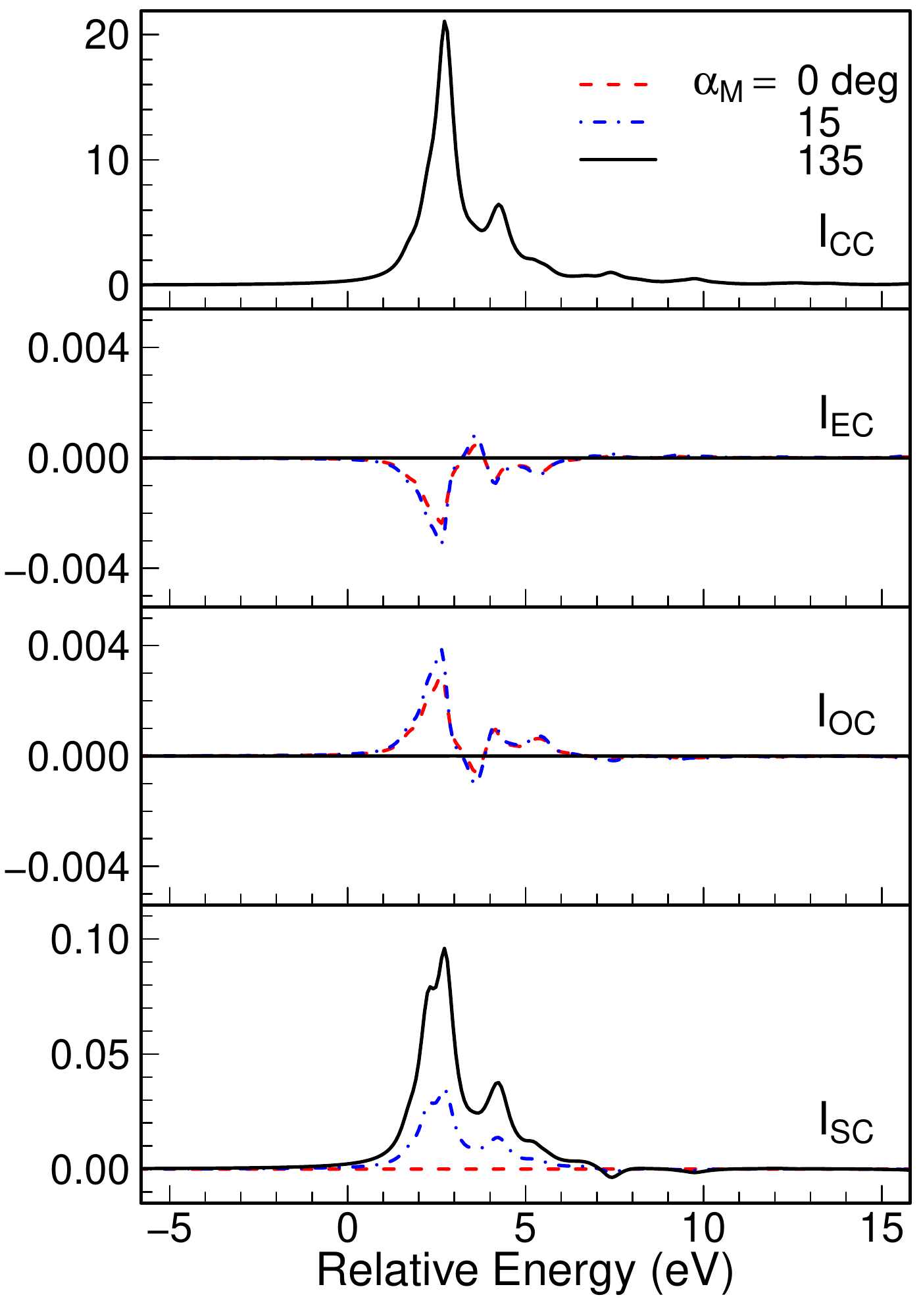}
\par\end{centering}
\caption{(Color online) The intensity $I_{\mathrm{CC}}$ and the MCD components
$\mathrm{Im}I_{\mathrm{EC}}$, $\mathrm{Im}I_{\mathrm{OC}}$, and
$\mathrm{Im}I_{\mathrm{SC}}$ as a function of relative energy of
the final states at $\alpha_{\mathrm{M}}=0^{\circ}$ (dashed line),
$15^{\circ}$ (dot-dash line) and $135^{\circ}$ (solid line) for
the $M_{1}$ edge XRS. The life time broadening is assumed to be $\Gamma=0.14$
eV. \label{fig:FeM_MCD-3s}}
\end{figure}

The MCD signal would be observed even at the $M_{1}$ edge with the
magnitude comparable to the $M_{\mathrm{2,3}}$ edge, because $I_{\mathrm{SC}}$
reflects the 3d spin polarization through the interaction $\boldsymbol{h}\cdot\boldsymbol{\sigma}$
between the electron spin and the radiation field in Eq. (\ref{eq:fs}).
The quadrupole transition, in which the factors $\tilde{j}_{2}\left(Q\right)$,
$\tilde{g}_{1}\left(Q\right)$, and $\tilde{f}_{2}\left(Q\right)$
are relevant, dominates the excitation process (3s$\rightarrow$3d)
at the $M_{1}$ edge. Due to the absence of the SOC in the 3s orbital,
$\mathrm{Im}I_{\mathrm{EC}}$ and $\mathrm{Im}I_{\mathrm{OC}}$ are
supposed to be small; those signals just reflect the 3d orbital polarization
due to the SOC in the 3d states. It is expected that $\mathrm{Im}I_{\mathrm{SC}}$
would be as large as that in the $M_{2,3}$ edge. Thus, the MCD signal
caused by only the 3d spin polarization would be observed.

In figure \ref{fig:FeM_MCD-3s}, the calculated $M_{1}$ edge XRS
and the MCD spectra are shown. The relative magnitude of the MCD signal
to the total intensity is in the same order with that in the $M_{2,3}$
edge spectra at the angle $\alpha_{\mathrm{M}}=135^{\circ}$. At this
angle $\alpha_{\mathrm{M}}$, the MCD components $\mathrm{Im}I_{\mathrm{EC}}$
and $\mathrm{Im}I_{\mathrm{OC}}$ are suppressed, so that the component
$\mathrm{Im}I_{\mathrm{SC}}$ alone contributes to the MCD signal.
Therefore, the MCD signal reflects only the spin moment of the 3d
orbital with the orbital magnetic quantum number $m=0$. At the angle
$\alpha_{\mathrm{M}}=0^{\circ}$, the component $\mathrm{Im}I_{\mathrm{SC}}$
is almost suppressed because the SOC in 3s orbital is absent and the
SOC in the 3d orbital is very small. The components $\mathrm{Im}I_{\mathrm{EC}}$
and $\mathrm{Im}I_{\mathrm{OC}}$ only weakly contribute to the MCD
signal due to the smallness of the SOC in 3d orbital. Because they
are very small and have opposite sign to each other, the total MCD
signal would be too small to be observed at the present stage.

Putting $M_{0}^{\prime}=-\sqrt{12\pi}\tilde{j}_{2}\left(Q\right)\left(Y_{20}\right)_{20,00},$
the integrated intensity $I_{\text{CC}}^{\sigma\sigma,\sigma\sigma}$
and component $\mathrm{Im}I_{\text{SC}}^{\sigma\pi,\sigma\sigma}$
at the angle $\alpha_{\mathrm{M}}=135^{\circ}$ might be given by

$\int_{E_{\mathrm{E}}}^{E_{\mathrm{C}}}\mathrm{d}xI_{\text{CC}}^{\sigma\sigma,\sigma\sigma}\left(x\right)=I_{0}M_{0}^{\prime2}\left(h_{0\uparrow}+h_{0\downarrow}\right)$,
and $\int_{E_{\mathrm{E}}}^{E_{\mathrm{C}}}\mathrm{d}x\mathrm{Im}I_{\text{SC}}^{\sigma\pi,\sigma\sigma}\left(x\right)=\frac{1}{2\sqrt{2}}\frac{\bar{E}}{m_{\text{e}}c^{2}}I_{0}M_{0}^{\prime2}\left(h_{0\uparrow}-h_{0\downarrow}\right)$,
respectively. Therefore, the ratio of the integrated MCD signals to
the integrated intensity could give the spin polarization ratio $\left(h_{0\uparrow}-h_{0\downarrow}\right)/\left(h_{0\uparrow}+h_{0\downarrow}\right)$
in the 3d orbitals with the magnetic quantum number $m=0$ as $C\int_{E_{\mathrm{E}}}^{E_{\mathrm{C}}}\mathrm{d}xI_{\mathrm{MCD}}\left(x\right)/\int_{E_{\mathrm{E}}}^{E_{\mathrm{C}}}\mathrm{d}xI_{\mathrm{TOT}}\left(x\right)$
with $C=\sqrt{2}\left(1+P_{3}\right)m_{\mathrm{e}}c^{2}/\left|P_{2}\right|\bar{E}$.

\section{Concluding remarks}

We investigated the XRS-MCD spectra by comparing the observed and
the theoretically calculated spectra at the $L_{2,3}$ and $M_{2,3}$
edges of ferromagnetic iron. We used the configuration interaction
calculation on the Anderson impurity model as a makeshift to simulate
the electronic structure of iron at the scattering center. The calculation
reproduced the observed spectra rather well in spite of the awkward
approximation for the strongly itinerant system. For more detailed
analysis, we would need a more sophisticated approximation and a model
which could appropriately reproduce the multiplet structure in the
excited state with taking into account both of the localized and itinerant
nature of the 3d electrons in the ferromagnetic iron. For the localized
electronic systems, the model used here may give more plausible results.

The MCD signals consist of the three components $\mathrm{Im}I_{\mathrm{EC}}$,
$\mathrm{Im}I_{\mathrm{OC}}$, and $\mathrm{Im}I_{\mathrm{SC}}$.
Their angle $\alpha_{\mathrm{M}}$ dependences of them are different.
Particularly, at $\alpha_{\mathrm{M}}=135^{\circ}$ in the right angle
scattering condition, $\mathrm{Im}I_{\mathrm{EC}}$ and $\mathrm{Im}I_{\mathrm{OC}}$
are suppressed if the total $J_{z}$ around the scattering site is
conserved or in the situation where the powder approximation is proper.
At this scattering geometry, the orbital magnetic quantum number $m$
is conserved both in the C- and S- transitions. The intensity $I_{\mathrm{CC}}$
is proportional to the 3d hole number, while the MCD component $\mathrm{Im}I_{\mathrm{SC}}$
is proportional to the difference of the number of the up and down
3d holes. Therefore, the information of the spin polarization in the
3d orbitals with the magnetic quantum numbers $m=0,\pm1$ may be obtained.

Here, we demonstrate the XRS-MCD spin sum rule at $\alpha_{\mathrm{M}}=135^{\circ}$.
The ratios of the integrated MCD signal and the total signal in the
observation $\int_{E_{\mathrm{E}}}^{E_{\mathrm{C}}}I_{\mathrm{MCD}}\left(x\right)\mathrm{d}x/\int_{E_{\mathrm{E}}}^{E_{\mathrm{C}}}I_{\mathrm{TOT}}\left(x\right)\mathrm{d}x$
is estimated to be $0.025\sim0.031$ at the $L_{2,3}$ edge with $E_{\mathrm{E}}=700\sim705$
and $E_{\mathrm{C}}=730\sim740$ eV. The ratio at the $M_{2,3}$ edge
is estimated to be $0.024\sim0.029$ with $E_{\mathrm{E}}=45\sim51$
and $E_{\mathrm{C}}=70\sim80$ eV. Using Eq. (\ref{eq:S/N}), these
ratios lead to the spin polarization ratio $S_{1}/N_{1}$ as $0.59\sim0.73$
for the $\mathrm{L_{2,3}}$-edge assuming $\bar{E}=10.2$ keV and
$0.58\sim0.70$ for the $\mathrm{M}_{2,3}$-edge assuming $\bar{E}=9.9$
keV. The value $S_{1}/N_{1}$ obtained by the CI calculation is $0.744$
for both the $\mathrm{L_{2.3}}$- and $\mathrm{M}_{2.3}$-edges. Assuming
that the 3d states accommodate $3.0$ holes per an iron atom, that
$h_{0\uparrow\left(\downarrow\right)}$, $\left(h_{1\uparrow\left(\downarrow\right)}+h_{1\uparrow\left(\downarrow\right)}\right)/2$,
and $\left(h_{2\uparrow\left(\downarrow\right)}+h_{2\uparrow\left(\downarrow\right)}\right)/2$
equal to each other, the local spin moment is estimated to be $1.7\sim2.3$$\mu_{\mathrm{B}}$.
The estimated value of the spin moment has large ambiguity at present
mainly due to smallness of the signal accumulation, we hope that the
difficulties in XRS-MCD experiment will be o\textbackslash{} in future
with the progress of the instrumentation.

We also demonstrated the XRS-MCD at the $M_{1}$ edge. Because the
MCD components $\mathrm{Im}I_{\mathrm{EC}}$ and $\mathrm{Im}I_{\mathrm{OC}}$
are mainly caused by the SOC in the core state, they are almost suppressed
and only weakly induced by the SOC in the 3d state. On the other hand,
the magnitude of the MCD component $\mathrm{Im}I_{\mathrm{SC}}$ is
comparable to that for $M_{2,3}$ edge because it reflects the spin
polarization in the 3d state. At the angle $\alpha_{\mathrm{M}}=135^{\circ}$,
it reflects the spin polarization in the 3d state with the magnetic
quantum number $m=0$. Therefore, the information of the spin polarization
in the 3d orbitals with the magnetic quantum numbers $m=0$ can be
obtained. By analyzing the MCD spectra at the $\mathrm{M}_{1}$-edge
together with the $M_{2,3}$ edge, it might be possible to obtain
the orbital resolved spin polarization. We have not yet known such
a simple procedure to obtain the information on the orbital moment
so far.

It is well known that the application of the spin sum rule in the
XAS-MCD requires careful consideration.\citep{Teramura1996,Teramura_1996}
Contrasting to the XAS-MCD, the sum rules (\ref{eq:Int_ICC}) and
(\ref{eq:Int_ISC}) do not subject to such a restriction. At angle
$\alpha_{\mathrm{M}}=135^{\circ}$, the transition processes leading
to the MCD component $\mathrm{Im}I_{\mathrm{SC}}$ and the intensity
$I_{\mathrm{CC}}$ are almost equivalent. Every final state due to
the C-transition and the S-transition coincide. In the S-transition,
the sign of the scattering amplitude is determined by the spin magnetic
quantum number of the excited electron. Thus, it is expected that
any decay processes result in the same effect on the spectral shape
of the total XRS intensity and the MCD signal. Therefore, analyzing
the total intensity and the MCD signal, we would be able to obtain
the information of the spin polarization in the 3d state. If we exploit
the $M_{1}$, $M_{2,3}$, and $M_{4,5}$ excitations to investigate
the 4d states, the orbital decomposed ($\left|m\right|=0,1,2$) information
about the spin polarization could be obtained. At angle $\alpha_{\mathrm{M}}=135^{\circ}$,
the total intensity and the MCD signal would show a quite similar
spectral curves to each other for the complete ferromagnetic state.
For the incomplete ferromagnetic state, these might show different
spectral curves. The spin resolved spectral curves might be obtained
by analyzing the total intensity and the MCD signal. We hope the XRS-MCD
will become one of useful tools to investigate the spin polarization
of the magnetic ions such as the XMD and the MCS.

\subsection*{Acknowledgment}

Author M.T. thanks Arata Tanaka for his kindness to allow us to use
the Xtls code adapted to our calculation and fruitful discussions.
The experiment was performed at BL12XU/SPring-8 with approvals of
SPring-8 and National Synchrotron Radiation Research Center, Taiwan
(Proposal No. 2016B4252/2016-2-042).

\bibliographystyle{apsrev4-1}
\bibliography{XRD_MCD}

\begin{thebibliography}{43}%
\makeatletter
\providecommand \@ifxundefined [1]{%
 \@ifx{#1\undefined}
}%
\providecommand \@ifnum [1]{%
 \ifnum #1\expandafter \@firstoftwo
 \else \expandafter \@secondoftwo
 \fi
}%
\providecommand \@ifx [1]{%
 \ifx #1\expandafter \@firstoftwo
 \else \expandafter \@secondoftwo
 \fi
}%
\providecommand \natexlab [1]{#1}%
\providecommand \enquote  [1]{``#1''}%
\providecommand \bibnamefont  [1]{#1}%
\providecommand \bibfnamefont [1]{#1}%
\providecommand \citenamefont [1]{#1}%
\providecommand \href@noop [0]{\@secondoftwo}%
\providecommand \href [0]{\begingroup \@sanitize@url \@href}%
\providecommand \@href[1]{\@@startlink{#1}\@@href}%
\providecommand \@@href[1]{\endgroup#1\@@endlink}%
\providecommand \@sanitize@url [0]{\catcode `\\12\catcode `\$12\catcode
  `\&12\catcode `\#12\catcode `\^12\catcode `\_12\catcode `\%12\relax}%
\providecommand \@@startlink[1]{}%
\providecommand \@@endlink[0]{}%
\providecommand \url  [0]{\begingroup\@sanitize@url \@url }%
\providecommand \@url [1]{\endgroup\@href {#1}{\urlprefix }}%
\providecommand \urlprefix  [0]{URL }%
\providecommand \Eprint [0]{\href }%
\providecommand \doibase [0]{http://dx.doi.org/}%
\providecommand \selectlanguage [0]{\@gobble}%
\providecommand \bibinfo  [0]{\@secondoftwo}%
\providecommand \bibfield  [0]{\@secondoftwo}%
\providecommand \translation [1]{[#1]}%
\providecommand \BibitemOpen [0]{}%
\providecommand \bibitemStop [0]{}%
\providecommand \bibitemNoStop [0]{.\EOS\space}%
\providecommand \EOS [0]{\spacefactor3000\relax}%
\providecommand \BibitemShut  [1]{\csname bibitem#1\endcsname}%
\let\auto@bib@innerbib\@empty
%</preamble>
\bibitem [{\citenamefont {Carra}\ \emph {et~al.}(1993)\citenamefont {Carra},
  \citenamefont {Thole}, \citenamefont {Altarelli},\ and\ \citenamefont
  {Wang}}]{Carra_1993}%
  \BibitemOpen
  \bibfield  {author} {\bibinfo {author} {\bibfnamefont {P.}~\bibnamefont
  {Carra}}, \bibinfo {author} {\bibfnamefont {B.~T.}\ \bibnamefont {Thole}},
  \bibinfo {author} {\bibfnamefont {M.}~\bibnamefont {Altarelli}}, \ and\
  \bibinfo {author} {\bibfnamefont {X.}~\bibnamefont {Wang}},\ }\href {\doibase
  10.1103/PhysRevLett.70.694} {\bibfield  {journal} {\bibinfo  {journal}
  {Physical Review Letters}\ }\textbf {\bibinfo {volume} {70}},\ \bibinfo
  {pages} {694} (\bibinfo {year} {1993})}\BibitemShut {NoStop}%
\bibitem [{\citenamefont {Thole}\ \emph {et~al.}(1992)\citenamefont {Thole},
  \citenamefont {Carra}, \citenamefont {Sette},\ and\ \citenamefont {van~der
  Laan}}]{Thole_1992}%
  \BibitemOpen
  \bibfield  {author} {\bibinfo {author} {\bibfnamefont {B.~T.}\ \bibnamefont
  {Thole}}, \bibinfo {author} {\bibfnamefont {P.}~\bibnamefont {Carra}},
  \bibinfo {author} {\bibfnamefont {F.}~\bibnamefont {Sette}}, \ and\ \bibinfo
  {author} {\bibfnamefont {G.}~\bibnamefont {van~der Laan}},\ }\href {\doibase
  10.1103/PhysRevLett.68.1943} {\bibfield  {journal} {\bibinfo  {journal}
  {Physical Review Letters}\ }\textbf {\bibinfo {volume} {68}},\ \bibinfo
  {pages} {1943} (\bibinfo {year} {1992})}\BibitemShut {NoStop}%
\bibitem [{\citenamefont {Nakamura}\ and\ \citenamefont
  {Suzuki}(2013)}]{Nakamura_2013}%
  \BibitemOpen
  \bibfield  {author} {\bibinfo {author} {\bibfnamefont {T.}~\bibnamefont
  {Nakamura}}\ and\ \bibinfo {author} {\bibfnamefont {M.}~\bibnamefont
  {Suzuki}},\ }\href {\doibase 10.7566/JPSJ.82.021006} {\bibfield  {journal}
  {\bibinfo  {journal} {Journal of the Physical Society of Japan}\ }\textbf
  {\bibinfo {volume} {82}},\ \bibinfo {pages} {021006} (\bibinfo {year}
  {2013})}\BibitemShut {NoStop}%
\bibitem [{\citenamefont {Kotani}\ and\ \citenamefont
  {Shin}(2001)}]{Kotani_2001}%
  \BibitemOpen
  \bibfield  {author} {\bibinfo {author} {\bibfnamefont {A.}~\bibnamefont
  {Kotani}}\ and\ \bibinfo {author} {\bibfnamefont {S.}~\bibnamefont {Shin}},\
  }\href {\doibase 10.1103/RevModPhys.73.203} {\bibfield  {journal} {\bibinfo
  {journal} {Reviews of Modern Physics}\ }\textbf {\bibinfo {volume} {73}},\
  \bibinfo {pages} {203} (\bibinfo {year} {2001})}\BibitemShut {NoStop}%
\bibitem [{\citenamefont {Hiraoka}\ \emph {et~al.}(2015)\citenamefont
  {Hiraoka}, \citenamefont {Takahashi}, \citenamefont {Wu}, \citenamefont
  {Lai}, \citenamefont {Tsuei},\ and\ \citenamefont {Huang}}]{Hiraoka_2015}%
  \BibitemOpen
  \bibfield  {author} {\bibinfo {author} {\bibfnamefont {N.}~\bibnamefont
  {Hiraoka}}, \bibinfo {author} {\bibfnamefont {M.}~\bibnamefont {Takahashi}},
  \bibinfo {author} {\bibfnamefont {W.~B.}\ \bibnamefont {Wu}}, \bibinfo
  {author} {\bibfnamefont {C.~H.}\ \bibnamefont {Lai}}, \bibinfo {author}
  {\bibfnamefont {K.~D.}\ \bibnamefont {Tsuei}}, \ and\ \bibinfo {author}
  {\bibfnamefont {D.~J.}\ \bibnamefont {Huang}},\ }\href {\doibase
  10.1103/PhysRevB.91.241112} {\bibfield  {journal} {\bibinfo  {journal}
  {Physical Review B}\ }\textbf {\bibinfo {volume} {91}},\ \bibinfo {pages}
  {241112(R)} (\bibinfo {year} {2015})}\BibitemShut {NoStop}%
\bibitem [{\citenamefont {Takahashi}\ and\ \citenamefont
  {Hiraoka}(2015)}]{Takahashi_2015}%
  \BibitemOpen
  \bibfield  {author} {\bibinfo {author} {\bibfnamefont {M.}~\bibnamefont
  {Takahashi}}\ and\ \bibinfo {author} {\bibfnamefont {N.}~\bibnamefont
  {Hiraoka}},\ }\href {\doibase 10.1103/PhysRevB.92.094441} {\bibfield
  {journal} {\bibinfo  {journal} {Physical Review B}\ }\textbf {\bibinfo
  {volume} {92}},\ \bibinfo {pages} {094441} (\bibinfo {year}
  {2015})}\BibitemShut {NoStop}%
\bibitem [{\citenamefont {Rueff}\ and\ \citenamefont
  {Shukla}(2010)}]{Rueff_2010}%
  \BibitemOpen
  \bibfield  {author} {\bibinfo {author} {\bibfnamefont {J.-P.}\ \bibnamefont
  {Rueff}}\ and\ \bibinfo {author} {\bibfnamefont {A.}~\bibnamefont {Shukla}},\
  }\href {\doibase 10.1103/RevModPhys.82.847} {\bibfield  {journal} {\bibinfo
  {journal} {Reviews of Modern Physics}\ }\textbf {\bibinfo {volume} {82}},\
  \bibinfo {pages} {847} (\bibinfo {year} {2010})}\BibitemShut {NoStop}%
\bibitem [{\citenamefont {Sternemann}\ and\ \citenamefont
  {Wilke}(2016)}]{Sternemann_2016}%
  \BibitemOpen
  \bibfield  {author} {\bibinfo {author} {\bibfnamefont {C.}~\bibnamefont
  {Sternemann}}\ and\ \bibinfo {author} {\bibfnamefont {M.}~\bibnamefont
  {Wilke}},\ }\href {\doibase 10.1080/08957959.2016.1198903} {\bibfield
  {journal} {\bibinfo  {journal} {High Pressure Research}\ }\textbf {\bibinfo
  {volume} {36}},\ \bibinfo {pages} {275} (\bibinfo {year} {2016})}\BibitemShut
  {NoStop}%
\bibitem [{\citenamefont {Nyrow}\ \emph
  {et~al.}(2014{\natexlab{a}})\citenamefont {Nyrow}, \citenamefont
  {Sternemann}, \citenamefont {Wilke}, \citenamefont {Gordon}, \citenamefont
  {Mende}, \citenamefont {Yava{\c{s}}}, \citenamefont {Simonelli},
  \citenamefont {Hiraoka}, \citenamefont {Sahle}, \citenamefont {Huotari},
  \citenamefont {Andreozzi}, \citenamefont {Woodland}, \citenamefont {Tolan},\
  and\ \citenamefont {Tse}}]{Nyrow_2014}%
  \BibitemOpen
  \bibfield  {author} {\bibinfo {author} {\bibfnamefont {A.}~\bibnamefont
  {Nyrow}}, \bibinfo {author} {\bibfnamefont {C.}~\bibnamefont {Sternemann}},
  \bibinfo {author} {\bibfnamefont {M.}~\bibnamefont {Wilke}}, \bibinfo
  {author} {\bibfnamefont {R.~A.}\ \bibnamefont {Gordon}}, \bibinfo {author}
  {\bibfnamefont {K.}~\bibnamefont {Mende}}, \bibinfo {author} {\bibfnamefont
  {H.}~\bibnamefont {Yava{\c{s}}}}, \bibinfo {author} {\bibfnamefont
  {L.}~\bibnamefont {Simonelli}}, \bibinfo {author} {\bibfnamefont
  {N.}~\bibnamefont {Hiraoka}}, \bibinfo {author} {\bibfnamefont {C.~J.}\
  \bibnamefont {Sahle}}, \bibinfo {author} {\bibfnamefont {S.}~\bibnamefont
  {Huotari}}, \bibinfo {author} {\bibfnamefont {G.~B.}\ \bibnamefont
  {Andreozzi}}, \bibinfo {author} {\bibfnamefont {A.~B.}\ \bibnamefont
  {Woodland}}, \bibinfo {author} {\bibfnamefont {M.}~\bibnamefont {Tolan}}, \
  and\ \bibinfo {author} {\bibfnamefont {J.~S.}\ \bibnamefont {Tse}},\ }\href
  {\doibase 10.1007/s00410-014-1012-8} {\bibfield  {journal} {\bibinfo
  {journal} {Contributions to Mineralogy and Petrology}\ }\textbf {\bibinfo
  {volume} {167}},\ \bibinfo {pages} {1012} (\bibinfo {year}
  {2014}{\natexlab{a}})}\BibitemShut {NoStop}%
\bibitem [{\citenamefont {Nyrow}\ \emph
  {et~al.}(2014{\natexlab{b}})\citenamefont {Nyrow}, \citenamefont {Tse},
  \citenamefont {Hiraoka}, \citenamefont {Desgreniers}, \citenamefont {Buning},
  \citenamefont {Mende}, \citenamefont {Tolan}, \citenamefont {Wilke},\ and\
  \citenamefont {Sternemann}}]{Nyrow_2014_2}%
  \BibitemOpen
  \bibfield  {author} {\bibinfo {author} {\bibfnamefont {A.}~\bibnamefont
  {Nyrow}}, \bibinfo {author} {\bibfnamefont {J.~S.}\ \bibnamefont {Tse}},
  \bibinfo {author} {\bibfnamefont {N.}~\bibnamefont {Hiraoka}}, \bibinfo
  {author} {\bibfnamefont {S.}~\bibnamefont {Desgreniers}}, \bibinfo {author}
  {\bibfnamefont {T.}~\bibnamefont {Buning}}, \bibinfo {author} {\bibfnamefont
  {K.}~\bibnamefont {Mende}}, \bibinfo {author} {\bibfnamefont
  {M.}~\bibnamefont {Tolan}}, \bibinfo {author} {\bibfnamefont
  {M.}~\bibnamefont {Wilke}}, \ and\ \bibinfo {author} {\bibfnamefont
  {C.}~\bibnamefont {Sternemann}},\ }\href {\doibase 10.1063/1.4886971}
  {\bibfield  {journal} {\bibinfo  {journal} {Applied Physics Letters}\
  }\textbf {\bibinfo {volume} {104}},\ \bibinfo {pages} {262408} (\bibinfo
  {year} {2014}{\natexlab{b}})}\BibitemShut {NoStop}%
\bibitem [{\citenamefont {van~der Laan}(2012)}]{van_der_Laan_2012}%
  \BibitemOpen
  \bibfield  {author} {\bibinfo {author} {\bibfnamefont {G.}~\bibnamefont
  {van~der Laan}},\ }\href {\doibase 10.1103/PhysRevB.86.035138} {\bibfield
  {journal} {\bibinfo  {journal} {Physical Review B}\ }\textbf {\bibinfo
  {volume} {86}},\ \bibinfo {pages} {035138} (\bibinfo {year}
  {2012})}\BibitemShut {NoStop}%
\bibitem [{\citenamefont {Huotari}\ \emph {et~al.}(2015)\citenamefont
  {Huotari}, \citenamefont {Suljoti}, \citenamefont {Sahle}, \citenamefont
  {Radel}, \citenamefont {Monaco},\ and\ \citenamefont
  {de~Groot}}]{Huotari_2015}%
  \BibitemOpen
  \bibfield  {author} {\bibinfo {author} {\bibfnamefont {S.}~\bibnamefont
  {Huotari}}, \bibinfo {author} {\bibfnamefont {E.}~\bibnamefont {Suljoti}},
  \bibinfo {author} {\bibfnamefont {C.~J.}\ \bibnamefont {Sahle}}, \bibinfo
  {author} {\bibfnamefont {S.}~\bibnamefont {Radel}}, \bibinfo {author}
  {\bibfnamefont {G.}~\bibnamefont {Monaco}}, \ and\ \bibinfo {author}
  {\bibfnamefont {F.~M.~F.}\ \bibnamefont {de~Groot}},\ }\href {\doibase
  10.1088/1367-2630/17/4/043041} {\bibfield  {journal} {\bibinfo  {journal}
  {New Journal of Physics}\ }\textbf {\bibinfo {volume} {17}},\ \bibinfo
  {pages} {043041} (\bibinfo {year} {2015})}\BibitemShut {NoStop}%
\bibitem [{\citenamefont {Teramura}\ \emph
  {et~al.}(1996{\natexlab{a}})\citenamefont {Teramura}, \citenamefont
  {Tanaka},\ and\ \citenamefont {Jo}}]{Teramura1996}%
  \BibitemOpen
  \bibfield  {author} {\bibinfo {author} {\bibfnamefont {Y.}~\bibnamefont
  {Teramura}}, \bibinfo {author} {\bibfnamefont {A.}~\bibnamefont {Tanaka}}, \
  and\ \bibinfo {author} {\bibfnamefont {T.}~\bibnamefont {Jo}},\ }\href
  {\doibase 10.1143/JPSJ.65.1053} {\bibfield  {journal} {\bibinfo  {journal}
  {Journal of the Physical Society of Japan}\ }\textbf {\bibinfo {volume}
  {65}},\ \bibinfo {pages} {1053} (\bibinfo {year}
  {1996}{\natexlab{a}})}\BibitemShut {NoStop}%
\bibitem [{\citenamefont {Teramura}\ \emph
  {et~al.}(1996{\natexlab{b}})\citenamefont {Teramura}, \citenamefont {Tanaka},
  \citenamefont {Thole},\ and\ \citenamefont {Jo}}]{Teramura_1996}%
  \BibitemOpen
  \bibfield  {author} {\bibinfo {author} {\bibfnamefont {Y.}~\bibnamefont
  {Teramura}}, \bibinfo {author} {\bibfnamefont {A.}~\bibnamefont {Tanaka}},
  \bibinfo {author} {\bibfnamefont {B.~T.}\ \bibnamefont {Thole}}, \ and\
  \bibinfo {author} {\bibfnamefont {T.}~\bibnamefont {Jo}},\ }\href {\doibase
  https://doi.org/10.1143/JPSJ.65.3056} {\bibfield  {journal} {\bibinfo
  {journal} {Journal of the Physical Society of Japan}\ }\textbf {\bibinfo
  {volume} {65}},\ \bibinfo {pages} {3056} (\bibinfo {year}
  {1996}{\natexlab{b}})}\BibitemShut {NoStop}%
\bibitem [{\citenamefont {Yoshida}\ and\ \citenamefont
  {Jo}(1991)}]{Yoshida_1991}%
  \BibitemOpen
  \bibfield  {author} {\bibinfo {author} {\bibfnamefont {A.}~\bibnamefont
  {Yoshida}}\ and\ \bibinfo {author} {\bibfnamefont {T.}~\bibnamefont {Jo}},\
  }\href {\doibase 10.1143/JPSJ.60.2098} {\bibfield  {journal} {\bibinfo
  {journal} {Journal of the Physical Society of Japan}\ }\textbf {\bibinfo
  {volume} {60}},\ \bibinfo {pages} {2098} (\bibinfo {year}
  {1991})}\BibitemShut {NoStop}%
\bibitem [{\citenamefont {Koide}\ \emph {et~al.}(1991)\citenamefont {Koide},
  \citenamefont {Shidara}, \citenamefont {Fukutani}, \citenamefont {Yamaguchi},
  \citenamefont {Fujimori},\ and\ \citenamefont {Kimura}}]{Koide_1991}%
  \BibitemOpen
  \bibfield  {author} {\bibinfo {author} {\bibfnamefont {T.}~\bibnamefont
  {Koide}}, \bibinfo {author} {\bibfnamefont {T.}~\bibnamefont {Shidara}},
  \bibinfo {author} {\bibfnamefont {H.}~\bibnamefont {Fukutani}}, \bibinfo
  {author} {\bibfnamefont {K.}~\bibnamefont {Yamaguchi}}, \bibinfo {author}
  {\bibfnamefont {A.}~\bibnamefont {Fujimori}}, \ and\ \bibinfo {author}
  {\bibfnamefont {S.}~\bibnamefont {Kimura}},\ }\href {\doibase
  10.1103/PhysRevB.44.4697} {\bibfield  {journal} {\bibinfo  {journal}
  {Physical Review B}\ }\textbf {\bibinfo {volume} {44}},\ \bibinfo {pages}
  {4697} (\bibinfo {year} {1991})}\BibitemShut {NoStop}%
\bibitem [{\citenamefont {Coster}\ and\ \citenamefont
  {Kronig}(1935)}]{Coster_1935}%
  \BibitemOpen
  \bibfield  {author} {\bibinfo {author} {\bibfnamefont {D.}~\bibnamefont
  {Coster}}\ and\ \bibinfo {author} {\bibfnamefont {R.~D.~L.}\ \bibnamefont
  {Kronig}},\ }\href {\doibase 10.1016/S0031-8914(35)90060-X} {\bibfield
  {journal} {\bibinfo  {journal} {Physica}\ }\textbf {\bibinfo {volume} {2}},\
  \bibinfo {pages} {13} (\bibinfo {year} {1935})}\BibitemShut {NoStop}%
\bibitem [{\citenamefont {Igarashi}\ and\ \citenamefont
  {Hirai}(1994)}]{Igarashi_1994}%
  \BibitemOpen
  \bibfield  {author} {\bibinfo {author} {\bibfnamefont {J.}~\bibnamefont
  {Igarashi}}\ and\ \bibinfo {author} {\bibfnamefont {K.}~\bibnamefont
  {Hirai}},\ }\href {\doibase 10.1103/PhysRevB.50.17820} {\bibfield  {journal}
  {\bibinfo  {journal} {Physical Review B}\ }\textbf {\bibinfo {volume} {50}},\
  \bibinfo {pages} {17820} (\bibinfo {year} {1994})}\BibitemShut {NoStop}%
\bibitem [{\citenamefont {Igarashi}\ and\ \citenamefont
  {Hirai}(1996)}]{Igarashi_1996}%
  \BibitemOpen
  \bibfield  {author} {\bibinfo {author} {\bibfnamefont {J.}~\bibnamefont
  {Igarashi}}\ and\ \bibinfo {author} {\bibfnamefont {K.}~\bibnamefont
  {Hirai}},\ }\href {\doibase 10.1103/PhysRevB.53.6442} {\bibfield  {journal}
  {\bibinfo  {journal} {Physical Review B}\ }\textbf {\bibinfo {volume} {53}},\
  \bibinfo {pages} {6442} (\bibinfo {year} {1996})}\BibitemShut {NoStop}%
\bibitem [{\citenamefont {Brouder}\ \emph {et~al.}(1996)\citenamefont
  {Brouder}, \citenamefont {Alouani},\ and\ \citenamefont
  {Bennemann}}]{Brouder_1996}%
  \BibitemOpen
  \bibfield  {author} {\bibinfo {author} {\bibfnamefont {C.}~\bibnamefont
  {Brouder}}, \bibinfo {author} {\bibfnamefont {M.}~\bibnamefont {Alouani}}, \
  and\ \bibinfo {author} {\bibfnamefont {K.~H.}\ \bibnamefont {Bennemann}},\
  }\href {\doibase 10.1103/PhysRevB.54.7334} {\bibfield  {journal} {\bibinfo
  {journal} {Physical Review B}\ }\textbf {\bibinfo {volume} {54}},\ \bibinfo
  {pages} {7334} (\bibinfo {year} {1996})}\BibitemShut {NoStop}%
\bibitem [{\citenamefont {Thole}\ and\ \citenamefont {van~der
  Laan}(1993)}]{Thole_1993}%
  \BibitemOpen
  \bibfield  {author} {\bibinfo {author} {\bibfnamefont {B.~T.}\ \bibnamefont
  {Thole}}\ and\ \bibinfo {author} {\bibfnamefont {G.}~\bibnamefont {van~der
  Laan}},\ }\href {\doibase https://doi.org/10.1103/PhysRevLett.70.2499}
  {\bibfield  {journal} {\bibinfo  {journal} {Physical Review Letters}\
  }\textbf {\bibinfo {volume} {70}},\ \bibinfo {pages} {2499} (\bibinfo {year}
  {1993})}\BibitemShut {NoStop}%
\bibitem [{\citenamefont {van~der Laan}\ and\ \citenamefont
  {Thole}(1993)}]{van_der_Laan_1993}%
  \BibitemOpen
  \bibfield  {author} {\bibinfo {author} {\bibfnamefont {G.}~\bibnamefont
  {van~der Laan}}\ and\ \bibinfo {author} {\bibfnamefont {B.~T.}\ \bibnamefont
  {Thole}},\ }\href {\doibase 10.1103/PhysRevB.48.210} {\bibfield  {journal}
  {\bibinfo  {journal} {Physical Review B}\ }\textbf {\bibinfo {volume} {48}},\
  \bibinfo {pages} {210} (\bibinfo {year} {1993})}\BibitemShut {NoStop}%
\bibitem [{\citenamefont {Cooper}\ \emph {et~al.}(2004)\citenamefont {Cooper},
  \citenamefont {Mijnarends}, \citenamefont {Shiotani}, \citenamefont {Sakai},\
  and\ \citenamefont {Bansil}}]{Cooper2004}%
  \BibitemOpen
  \bibfield  {author} {\bibinfo {author} {\bibfnamefont {M.}~\bibnamefont
  {Cooper}}, \bibinfo {author} {\bibfnamefont {P.}~\bibnamefont {Mijnarends}},
  \bibinfo {author} {\bibfnamefont {N.}~\bibnamefont {Shiotani}}, \bibinfo
  {author} {\bibfnamefont {N.}~\bibnamefont {Sakai}}, \ and\ \bibinfo {author}
  {\bibfnamefont {A.}~\bibnamefont {Bansil}},\ }\href {\doibase
  10.1093/acprof:oso/9780198501688.001.0001} {\emph {\bibinfo {title} {X-Ray
  Compton Scattering}}}\ (\bibinfo  {publisher} {Oxford University Press, New
  York},\ \bibinfo {year} {2004})\BibitemShut {NoStop}%
\bibitem [{\citenamefont {Blume}(1985)}]{Blume_1985}%
  \BibitemOpen
  \bibfield  {author} {\bibinfo {author} {\bibfnamefont {M.}~\bibnamefont
  {Blume}},\ }\href {\doibase 10.1063/1.335023} {\bibfield  {journal} {\bibinfo
   {journal} {Journal of Applied Physics}\ }\textbf {\bibinfo {volume} {57}},\
  \bibinfo {pages} {3615} (\bibinfo {year} {1985})}\BibitemShut {NoStop}%
\bibitem [{\citenamefont {Garvie}\ and\ \citenamefont
  {Buseck}(2004)}]{Garvie_2004}%
  \BibitemOpen
  \bibfield  {author} {\bibinfo {author} {\bibfnamefont {L.~A.}\ \bibnamefont
  {Garvie}}\ and\ \bibinfo {author} {\bibfnamefont {P.~R.}\ \bibnamefont
  {Buseck}},\ }\href {\doibase 10.2138/am-2004-0402} {\bibfield  {journal}
  {\bibinfo  {journal} {American Mineralogist}\ }\textbf {\bibinfo {volume}
  {89}},\ \bibinfo {pages} {485} (\bibinfo {year} {2004})}\BibitemShut
  {NoStop}%
\bibitem [{\citenamefont {Taguchi}\ \emph {et~al.}(1997)\citenamefont
  {Taguchi}, \citenamefont {Uozumi},\ and\ \citenamefont
  {Kotani}}]{Taguchi_1997}%
  \BibitemOpen
  \bibfield  {author} {\bibinfo {author} {\bibfnamefont {M.}~\bibnamefont
  {Taguchi}}, \bibinfo {author} {\bibfnamefont {T.}~\bibnamefont {Uozumi}}, \
  and\ \bibinfo {author} {\bibfnamefont {A.}~\bibnamefont {Kotani}},\ }\href
  {\doibase 10.1143/JPSJ.66.247} {\bibfield  {journal} {\bibinfo  {journal}
  {Journal of the Physical Society of Japan}\ }\textbf {\bibinfo {volume}
  {66}},\ \bibinfo {pages} {247} (\bibinfo {year} {1997})}\BibitemShut
  {NoStop}%
\bibitem [{\citenamefont {Inami}(2017)}]{Inami_2017}%
  \BibitemOpen
  \bibfield  {author} {\bibinfo {author} {\bibfnamefont {T.}~\bibnamefont
  {Inami}},\ }\href {\doibase 10.1103/PhysRevLett.119.137203} {\bibfield
  {journal} {\bibinfo  {journal} {Physical Review Letter}\ }\textbf {\bibinfo
  {volume} {119}},\ \bibinfo {pages} {137203} (\bibinfo {year}
  {2017})}\BibitemShut {NoStop}%
\bibitem [{\citenamefont {Fr\"ohlich}\ and\ \citenamefont
  {Studer}(1993)}]{Fr_hlich_1993}%
  \BibitemOpen
  \bibfield  {author} {\bibinfo {author} {\bibfnamefont {J.}~\bibnamefont
  {Fr\"ohlich}}\ and\ \bibinfo {author} {\bibfnamefont {U.~M.}\ \bibnamefont
  {Studer}},\ }\href {\doibase 10.1103/RevModPhys.65.733} {\bibfield  {journal}
  {\bibinfo  {journal} {Reviews of Modern Physics}\ }\textbf {\bibinfo {volume}
  {65}},\ \bibinfo {pages} {733} (\bibinfo {year} {1993})}\BibitemShut
  {NoStop}%
\bibitem [{\citenamefont {Trammell}(1953)}]{Trammell_1953}%
  \BibitemOpen
  \bibfield  {author} {\bibinfo {author} {\bibfnamefont {G.~T.}\ \bibnamefont
  {Trammell}},\ }\href {\doibase 10.1103/physrev.92.1387} {\bibfield  {journal}
  {\bibinfo  {journal} {Physical Review}\ }\textbf {\bibinfo {volume} {92}},\
  \bibinfo {pages} {1387} (\bibinfo {year} {1953})}\BibitemShut {NoStop}%
\bibitem [{\citenamefont {Berestetskii}\ \emph {et~al.}(1982)\citenamefont
  {Berestetskii}, \citenamefont {Lifshitz},\ and\ \citenamefont
  {Pitaevskii}}]{1982}%
  \BibitemOpen
  \bibfield  {author} {\bibinfo {author} {\bibfnamefont {V.~B.}\ \bibnamefont
  {Berestetskii}}, \bibinfo {author} {\bibfnamefont {E.~M.}\ \bibnamefont
  {Lifshitz}}, \ and\ \bibinfo {author} {\bibfnamefont {L.~P.}\ \bibnamefont
  {Pitaevskii}},\ }\href {\doibase 10.1016/B978-0-08-050346-2.50023-4} {\emph
  {\bibinfo {title} {Quantum Electrodynamics}}}\ (\bibinfo  {publisher}
  {Elsevier, New York},\ \bibinfo {year} {1982})\ p.~\bibinfo {pages}
  {ii}\BibitemShut {NoStop}%
\bibitem [{\citenamefont {Varshalovich}\ \emph {et~al.}(1988)\citenamefont
  {Varshalovich}, \citenamefont {Moskalev},\ and\ \citenamefont
  {Khersonskii}}]{Varshalovich_1988}%
  \BibitemOpen
  \bibfield  {author} {\bibinfo {author} {\bibfnamefont {D.~A.}\ \bibnamefont
  {Varshalovich}}, \bibinfo {author} {\bibfnamefont {A.~N.}\ \bibnamefont
  {Moskalev}}, \ and\ \bibinfo {author} {\bibfnamefont {V.~K.}\ \bibnamefont
  {Khersonskii}},\ }\href {\doibase 10.1142/0270} {\emph {\bibinfo {title}
  {Quantum Theory of Angular Momentum}}}\ (\bibinfo  {publisher} {World
  Scientific, Singapore},\ \bibinfo {year} {1988})\BibitemShut {NoStop}%
\bibitem [{\citenamefont {Blume}\ and\ \citenamefont
  {Gibbs}(1988)}]{Blume_1988}%
  \BibitemOpen
  \bibfield  {author} {\bibinfo {author} {\bibfnamefont {M.}~\bibnamefont
  {Blume}}\ and\ \bibinfo {author} {\bibfnamefont {D.}~\bibnamefont {Gibbs}},\
  }\href {\doibase 10.1103/PhysRevB.37.1779} {\bibfield  {journal} {\bibinfo
  {journal} {Physical Review B}\ }\textbf {\bibinfo {volume} {37}},\ \bibinfo
  {pages} {1779} (\bibinfo {year} {1988})}\BibitemShut {NoStop}%
\bibitem [{\citenamefont {Lovesey}(1987)}]{Lovesey_1987}%
  \BibitemOpen
  \bibfield  {author} {\bibinfo {author} {\bibfnamefont {S.~W.}\ \bibnamefont
  {Lovesey}},\ }\href {\doibase 10.1088/0022-3719/20/34/003} {\bibfield
  {journal} {\bibinfo  {journal} {Journal of Physics C: Solid State Physics}\
  }\textbf {\bibinfo {volume} {20}},\ \bibinfo {pages} {5625} (\bibinfo {year}
  {1987})}\BibitemShut {NoStop}%
\bibitem [{\citenamefont {Laundy}\ \emph {et~al.}(1991)\citenamefont {Laundy},
  \citenamefont {Collins},\ and\ \citenamefont {Rollason}}]{Laundy_1991}%
  \BibitemOpen
  \bibfield  {author} {\bibinfo {author} {\bibfnamefont {D.}~\bibnamefont
  {Laundy}}, \bibinfo {author} {\bibfnamefont {S.~P.}\ \bibnamefont {Collins}},
  \ and\ \bibinfo {author} {\bibfnamefont {A.~J.}\ \bibnamefont {Rollason}},\
  }\href {\doibase 10.1088/0953-8984/3/3/011} {\bibfield  {journal} {\bibinfo
  {journal} {Journal of Physics: Condensed Matter}\ }\textbf {\bibinfo {volume}
  {3}},\ \bibinfo {pages} {369} (\bibinfo {year} {1991})}\BibitemShut {NoStop}%
\bibitem [{\citenamefont {Jo}\ and\ \citenamefont {Sawatzky}(1991)}]{Jo_1991}%
  \BibitemOpen
  \bibfield  {author} {\bibinfo {author} {\bibfnamefont {T.}~\bibnamefont
  {Jo}}\ and\ \bibinfo {author} {\bibfnamefont {G.~A.}\ \bibnamefont
  {Sawatzky}},\ }\href {\doibase 10.1103/PhysRevB.43.8771} {\bibfield
  {journal} {\bibinfo  {journal} {Physical Review B}\ }\textbf {\bibinfo
  {volume} {43}},\ \bibinfo {pages} {8771} (\bibinfo {year}
  {1991})}\BibitemShut {NoStop}%
\bibitem [{\citenamefont {Tanaka}\ \emph {et~al.}(1992)\citenamefont {Tanaka},
  \citenamefont {Jo},\ and\ \citenamefont {Sawatzky}}]{Tanaka_1992}%
  \BibitemOpen
  \bibfield  {author} {\bibinfo {author} {\bibfnamefont {A.}~\bibnamefont
  {Tanaka}}, \bibinfo {author} {\bibfnamefont {T.}~\bibnamefont {Jo}}, \ and\
  \bibinfo {author} {\bibfnamefont {G.~A.}\ \bibnamefont {Sawatzky}},\ }\href
  {\doibase 10.1143/JPSJ.61.2636} {\bibfield  {journal} {\bibinfo  {journal}
  {Journal of the Physical Society of Japan}\ }\textbf {\bibinfo {volume}
  {61}},\ \bibinfo {pages} {2636} (\bibinfo {year} {1992})}\BibitemShut
  {NoStop}%
\bibitem [{Note1()}]{Note1}%
  \BibitemOpen
  \bibinfo {note} {We assumed $n_{0}=4$, which gives most plausible
  results.}\BibitemShut {Stop}%
\bibitem [{\citenamefont {Cowan}(1981)}]{cowan1981theory}%
  \BibitemOpen
  \bibfield  {author} {\bibinfo {author} {\bibfnamefont {R.}~\bibnamefont
  {Cowan}},\ }\href {https://books.google.co.jp/books?id=tHOXLrXkJRgC} {\emph
  {\bibinfo {title} {The Theory of Atomic Structure and Spectra}}},\ Los Alamos
  Series in Basic and Applied Sciences\ (\bibinfo  {publisher} {University of
  California Press, Berkeley},\ \bibinfo {year} {1981})\BibitemShut {NoStop}%
\bibitem [{\citenamefont {Santoni}\ and\ \citenamefont
  {Himpsel}(1991)}]{Santoni_1991}%
  \BibitemOpen
  \bibfield  {author} {\bibinfo {author} {\bibfnamefont {A.}~\bibnamefont
  {Santoni}}\ and\ \bibinfo {author} {\bibfnamefont {F.~J.}\ \bibnamefont
  {Himpsel}},\ }\href {\doibase 10.1103/PhysRevB.43.1305} {\bibfield  {journal}
  {\bibinfo  {journal} {Physical Review B}\ }\textbf {\bibinfo {volume} {43}},\
  \bibinfo {pages} {1305} (\bibinfo {year} {1991})}\BibitemShut {NoStop}%
\bibitem [{\citenamefont {Okada}\ \emph {et~al.}(1993)\citenamefont {Okada},
  \citenamefont {Kotani}, \citenamefont {Ogasawara}, \citenamefont {Seino},\
  and\ \citenamefont {Thole}}]{Okada_1993}%
  \BibitemOpen
  \bibfield  {author} {\bibinfo {author} {\bibfnamefont {K.}~\bibnamefont
  {Okada}}, \bibinfo {author} {\bibfnamefont {A.}~\bibnamefont {Kotani}},
  \bibinfo {author} {\bibfnamefont {H.}~\bibnamefont {Ogasawara}}, \bibinfo
  {author} {\bibfnamefont {Y.}~\bibnamefont {Seino}}, \ and\ \bibinfo {author}
  {\bibfnamefont {B.~T.}\ \bibnamefont {Thole}},\ }\href {\doibase
  10.1103/physrevb.47.6203} {\bibfield  {journal} {\bibinfo  {journal}
  {Physical Review B}\ }\textbf {\bibinfo {volume} {47}},\ \bibinfo {pages}
  {6203} (\bibinfo {year} {1993})}\BibitemShut {NoStop}%
\bibitem [{Note2()}]{Note2}%
  \BibitemOpen
  \bibinfo {note} {In the experiments at the L- and M-edges , the incident
  photon energy is scanned over a specific range to detect emitted photons with
  an energy of 9888 eV.}\BibitemShut {Stop}%
\bibitem [{\citenamefont {Doniach}\ \emph {et~al.}(1971)\citenamefont
  {Doniach}, \citenamefont {Platzman},\ and\ \citenamefont
  {Yue}}]{Doniach_1971}%
  \BibitemOpen
  \bibfield  {author} {\bibinfo {author} {\bibfnamefont {S.}~\bibnamefont
  {Doniach}}, \bibinfo {author} {\bibfnamefont {P.~M.}\ \bibnamefont
  {Platzman}}, \ and\ \bibinfo {author} {\bibfnamefont {J.~T.}\ \bibnamefont
  {Yue}},\ }\href {\doibase 10.1103/PhysRevB.4.3345} {\bibfield  {journal}
  {\bibinfo  {journal} {Physical Review B}\ }\textbf {\bibinfo {volume} {4}},\
  \bibinfo {pages} {3345} (\bibinfo {year} {1971})}\BibitemShut {NoStop}%
\bibitem [{\citenamefont {Nozi{\`{e}}res}\ and\ \citenamefont
  {Abrahams}(1974)}]{Nozi_res_1974}%
  \BibitemOpen
  \bibfield  {author} {\bibinfo {author} {\bibfnamefont {P.}~\bibnamefont
  {Nozi{\`{e}}res}}\ and\ \bibinfo {author} {\bibfnamefont {E.}~\bibnamefont
  {Abrahams}},\ }\href {\doibase 10.1103/PhysRevB.10.3099} {\bibfield
  {journal} {\bibinfo  {journal} {Physical Review B}\ }\textbf {\bibinfo
  {volume} {10}},\ \bibinfo {pages} {3099} (\bibinfo {year}
  {1974})}\BibitemShut {NoStop}%
\end{thebibliography}%

\end{document}